\documentclass[11pt,english,eqsecnum,english,prd,aps,nofootinbib,superscriptaddress,amsfonts,longbibliography,showpacs,showkeys, tightenlines]{revtex4-1}
\usepackage{mathptmx}
\usepackage[T1]{fontenc}
\usepackage[latin9]{inputenc}
\usepackage{babel}
\usepackage{verbatim}
\usepackage{mathrsfs}
\usepackage{amsmath}
\usepackage{amssymb}
\usepackage{cancel}
\usepackage{graphicx}
\usepackage[unicode=true,pdfusetitle,
 bookmarks=true,bookmarksnumbered=false,bookmarksopen=false,
 breaklinks=true,pdfborder={0 0 1},backref=false,colorlinks=false]
 {hyperref}

\makeatletter
\allowdisplaybreaks[1]

\makeatother

\begin{document}

\title{Path integral polymer propagator of relativistic and non-relativistic
particles}

\author{Hugo A. Morales-T\'{e}cotl}
\email{hugo@xanum.uam.mx}

\author{Saeed Rastgoo}
\email{saeed@xanum.uam.mx}

\author{Juan C. Ruelas}
\email{j.carlos.ruelas.v@gmail.com}

\affiliation{Departamento de F\'{i}sica, Universidad Aut\'{o}noma Metropolitana
- Iztapalapa~\\
 San Rafael Atlixco 186, Mexico D.F. 09340, Mexico}

\date{\today}
\begin{abstract}
A recent proposal to connect the loop quantization with the spin foam
model for cosmology via the path integral is hereby adapted to the
case of mechanical systems within the framework of the so called polymer
quantum mechanics. The mechanical models we consider are deparametrized
and thus the group averaging technique is used to deal with the corresponding
constraints. The transition amplitudes are written in a vertex expansion
form used in  the spin foam models, where here a vertex is actually
a jump in position. Polymer propagators previously obtained by spectral
methods for a nonrelativistic polymer particle, both free and in
a box, are regained with this method and as a new result we obtain the polymer propagator of the relativistic particle.
All of them reduce to their standard form in the continuum limit for which
the length scale parameter of the polymer quantization is taken to
be small. Our results are robust thanks to their analytic and exact character which in turn come from the fact that presented models are solvable. They lend support to the vertex expansion scheme of the polymer path integral explored before in a formal way for cosmological models. Some possible future developments are commented upon in the discussion. 
\end{abstract}


\maketitle

\section{Introduction\label{sec:Intro}}
The space-time singularities of General Relativity \citep{HE}
and the ultraviolet divergences in Quantum Field Theory (QFT) \citep{Weinberg}
are open physical problems deeply connected to the underlying structure
of space-time which is classically assumed to be a smooth manifold
with a metric obeying equations of general relativity. There have
been numerous attempts to understand and deal with these issues in
approaches like string theory \citep{Zweibach,GSW,Polchinsky}
and loop quantum Gravity (LQG) \citep{Bojowald,Thiemann,Rovelli}.

The need to cope with these divergences has led to investigation proposals
in quantum gravity that may alleviate such difficulties, possibly
by modifying the underlying space-time structure at the Planck scale.
More concretely, a common understanding coming out of many of these
attempts is that the high energy structure of the spacetime is fundamentally
discrete. As a salient candidate, LQG has been successful to resolve
the classical gravitational singularities of various cosmological
and black hole models (there is a huge body of research done in both of these topics; for a sample of historical and recent works see \cite{ASP, Ashtekar:0812, Ashtekar:0509, R.Gambini2013, Corichi1608} and the references within). This theory derives the discreteness of the
spacetime via the quantization of spacetimes degrees of freedom, the
so called holonomies and their conjugate fluxes. LQG comes in two
versions, a Hamiltonian one and a path integral version or the spin
foam model \citep{RovelliSF} which is based on the
transition amplitudes a la Feynman and thus a manifestly covariant
theory.

The representation  used in LQG is a departure from
the regular representation of the Weyl relations. Such a representation
is called polymer quantization. As we will see, this type of quantization
leads to a representation of the classical algebra that is not unitarily
equivalent to the Schr\"{o}dinger representation.
This is a consequence of the fact that the weak continuity assumption of the Stone-von Neumann theorem does not hold for such polymer representation. The lack of continuity, in turn, is the
result of a specific choice of topology (i.e. a discrete one). Polymer
representation exhibits two standard polarization in which either
configuration variables or the momenta are inherently discrete, and
the conjugate ones take values on a compact space ($S^1$
for example). 

Applying this type of quatization to mechanical system, i.e. systems
with finite number of degrees of freedom is known as polymer quantum
mechanics \citep{AFW}, while combining it with Feynman
path integral formulation leads to the polymer path integral formulation.
Also, if it is applied to each of the infinite modes of the fields
\citep{Hossain} it would yield a polymer field model.

Historically, combining path integration with polymer quantization
was considered originally in \cite{Husain-not-pub}
and more recently has been considered along different lines. For instance
Bianchi I cosmological model \cite{Liu:2012xp} and
its effective dynamics have been studied including the cosmological
constant \cite{Fujio:2012zz}. All the isotropic models
both in the deparametrized and timeless frameworks were worked out
in \cite{Huang:2011es}. Even alternative dynamics
has been used which however yields the same effective dynamics \cite{Qin:2012gaa}.
Similar techniques have been applied to consistent histories approach
in cosmology \cite{Craig:2012zz,Craig:2016fhk,Craig:2016iuw}.
Even a coherent state functional has been studied for some models
\cite{Qin:2011hx,Qin:2012xh}. Interestingly, a parallelism
between the two point functions for cosmology and the relativistic
particle has shed some light on the timeless and deparametrized frameworks
\cite{Calcagni:2010ad}. For other systems some mechanical
models have been considered that resemble the problematics in analyzing
the semiclassical approximation of certain black holes \cite{Hugo-Saeed-Daniel}.
Also polymer field theories have been dealt with by adopting this
approach \cite{Kajuri:2015oza,Kajuri:2014kva}. 

Recently, a detailed connection between the Hamiltonain LQG and the
spin foam model for the case of homogeneous isotropic cosmic models
has been developed \citep{AshtekarPI,A2} showing
that both Hamiltonian transition amplitudes, time deparametrized and
time reparametrization invariant, can be written in a vertex expansion
form, typical of the spin foam models, and they actually coincide.
Based on this, it is interesting to apply this method also to mechanical
systems with finite degrees of freedom (i.e. vertex expansion together
with polymer quantum mechanics) to see how are the results different
from normal Schr\"{o}dinger representation.
It is particularly useful since in these cases, the propagator can
be obtained analytically, exactly and explicitly due to solvable nature
of many of these systems. Furthermore, it provides a direct and exact
consistency check for the above method and also hopefully sheds more
light on the physics behind the idea. In this spirit, we try to derive the
polymer path integral of three mechanical systems, the free nonrelativistic
particle, the same particle in a box, and the relativistic particle. 

This work is organized as follows. In Sec. \ref{sec:PQM}
we present an introduction to polymer representation and its kinematics
and dynamics, for a reader that is not quite familiar with the subject.
In Sec. \ref{sec:poly-depar-prop} the tools that
are used in the rest of the work are introduced. These include, the
nonrelativistic and relativistic deparamatrization framework of constrained
systems, group averaging technique to deal with solving quantum constrains
and defining inner product on their representation space, and the
generic form of the polymer path integral of a deparametrized system.
Sec. \ref{sec:generic-amp-poly} is devoted to present
the method introduced in \citep{AshtekarPI,A2}
to cast loop quantum cosmology in spin foam framework. Here we develop
it for a single particle to be used in the following sections. In
Sec. \ref{sec:non-rel-partcl}, we use all the material
introduced in previous sections to derive the polymer path integral
of a nonrelativistic particle, both free and in a box. The continuum
limit of such a path integral is also derived. In Sec. \ref{sec:rel-partcl},
the polymer path integral of a relativistic particle and its continuum
limit are derived. This is a new result as far as the authors understand.
Finally in Sec. \ref{sec:discuss}, we summarize our
findings and make several remarks about the work.

\section{Polymer Quantum Mechanics\label{sec:PQM}}

\subsection{Weyl relation and its Schr\"{o}dinger representation}

An important part of any (canonical) quantization procedure is how
to represent the classical phase space variables and their algebra,
and consequently the functions of phase space, as linear operators
on a suitable Hilbert space. Usually the representations that are
used are unitarily equivalent to the Schr\"{o}dinger representation
as we will see. However not all of them are so, and the polymer representation
\citep{AFW} is an example of such inequivalent representation. In
order to understand the basic foundations of the polymer representation,
we will briefly look at the Weyl group and its representations and
then describe the basics of polymer representation of this group.

The classical Poisson algebra of a simple quantum mechanical system
between a pair of canonical variables can be written as $\{q,p\}=1$.
If we represent this on a Hilbert space such that 
\begin{equation}
[\hat{q},\hat{p}]=\hat{q}\hat{p}-\hat{p}\hat{q}=i\hat{1},\label{eq:qt-commut}
\end{equation}
it can be shown that $\hat{q}$ and $\hat{p}$ cannot both be bounded
operators\footnote{A linear operator (transformation) $L:V_{1}\rightarrow V_{2}$ from
a normed vector space $V_{1}$ to another one $V_{2}$ is a map for
which the ratio of the norm of $L(v)$ to that of $v$, for all $v\in V_{1}$
and $v\neq0$ is bounded: $\frac{\Vert L(v)\Vert}{\Vert v\Vert}<\infty,\,\forall v\in V_{1}\,\land\,v\neq0$.} \citep{Esposito}. This is called the Wintner theorem \citep{Wintner}.
Obviously, the right hand side of this relation is well-defined over
all of the Hilbert space. However, since in the left hand side, at
least one of the operators is not bounded, this side is not well-defined
unless we introduce further conditions on the operators to make it
so. One way is to specify the domain of the operators in Hilbert space
on which they are bounded, but this can get very complicated. A simpler
way around this problem was proposed by Weyl \citep{Weyl-1931} in
which one works with the ``exponential version'' of these operators.
Since these can be unitary operators and these types of operators
are bounded, then we need not to worry about the issue of boundedness.
Let us see this in more detail.

We first probe the classical theory and then move to the quantum regime.
Following the Weyl proposal, we can write classically 
\begin{equation}
U=e^{i\alpha q},\,\,\,\,\,\,\,\,V=e^{i\beta p}.
\end{equation}
which are classical unitary objects if $q$ and $p$ are real. Using
the Baker-Campbell-Hausdorff theorem, for a pair $(q,p)$ that obey
the canonical commutation relations $\{q,p\}=1$, we can write 
\begin{equation}
U(\alpha)V(\beta)=e^{-\alpha\beta\{q,p\}}V(\beta)U(\alpha).\label{eq:Weyl-classic}
\end{equation}
Note that this is strictly valid only for $U$ and $V$ that are generated
by classical canonical variables, but not for any general $U$ and
$V$ that are unitary.

Moving to the quantum version (i.e. the representation of the operators
on the Hilbert space), we can see that if both of the canonical pair
of operators $\hat{q}$ and $\hat{p}$ were well-defined on the Hilbert
space, we could have written their quantum version of exponentiation
\begin{equation}
\hat{U}=e^{i\alpha\hat{q}},\,\,\,\,\,\,\,\,\hat{V}=e^{i\beta\hat{p}}\label{eq:exp-q-p-qt}
\end{equation}
and (\ref{eq:Weyl-classic}) would have become (using the Dirac prescription
$[\hat{q},\hat{p}]=i\hbar\widehat{\{q,p\}}$) 
\begin{equation}
\hat{U}(\alpha)\hat{V}(\beta)=e^{\frac{i}{\hbar}\alpha\beta}\hat{V}(\beta)\hat{U}(\alpha).\label{eq:Weyl-main}
\end{equation}
This is called the Weyl relation. Now the Weyl prescription is that,
since not always both of generators $\hat{q},\hat{p}$ can be represented
as well-defined operators on the Hilbert space due to boundedness
issues (and also continuity, see below) and thus we can not always
literally exponentiate them to get their associated unitary operator\footnote{Or equivalently, cannot always expand $\hat{U}(\alpha)$ or $\hat{V}(\beta)$
as Taylor series in $\hat{q}$ or $\hat{p}$ respectively.}, we can just forget about (\ref{eq:qt-commut}) and instead take
the Weyl relations (\ref{eq:Weyl-main}) as fundamental relations,
and also take $\hat{U}(\alpha)$ and $\hat{V}(\beta)$ as the basic
operators of the theory on their own and not as exponentiation of
a generator. Then, if under some conditions, it is possible to write
$\hat{U}$ and $\hat{V}$ as exponentials of some generators as in
(\ref{eq:exp-q-p-qt}), we can easily regain (\ref{eq:qt-commut})
by using Baker-Campbell-Hausdorff theorem. These conditions, which
we will discuss briefly in the following, are the ones that distinguish
between the Schr\"{o}dinger and polymer representations.

Consider a one-parameter group of linear operators $\{L_{t}\}$ defined
on a Hilbert space $\mathscr{H}$, 
\begin{equation}
L_{t}:\mathbb{R}\rightarrow\textrm{End}(\mathscr{H})\,\,\,\,\textrm{and}\,\,\,\,\,\,\,\,\,L_{t}L_{s}=L_{t+s}\,\,\,\forall t,s\in\mathbb{R},\,\,\,\,\,\,L_{0}=1.
\end{equation}
This one-parameter group is called weakly continuous \citep{deOliviera}
if 
\begin{equation}
\lim_{t\rightarrow t_{0}}\langle\psi|L_{t}|\phi\rangle=\langle\psi|L_{t_{0}}|\phi\rangle,\,\,\,\,\,\,\,\forall|\psi\rangle,|\phi\rangle\in\mathscr{H},\,\,\forall t_{0}\in\mathbb{R}.
\end{equation}
It is called strongly continuous if 
\begin{equation}
\lim_{t\rightarrow t_{0}}L_{t}|\phi\rangle=L_{t_{0}}|\phi\rangle,\,\,\,\,\,\,\,\forall|\phi\rangle\in\mathscr{H},\,\,\forall t_{0}\in\mathbb{R}.
\end{equation}
In the case where $\{L_{t}\}$ is strongly continuous, there are states
for which the limit 
\begin{equation}
\lim_{t\rightarrow0}\frac{L_{t}|\psi\rangle-L_{0}|\psi\rangle}{t}=ig|\psi\rangle,\,\,\,\,\,\,|\psi\rangle\in\mathscr{H}
\end{equation}
exists, and so within this domain of states, one can define the infinitesimal
generators $g$ of the one-parameter group $\{L_{t}\}$ such that
\begin{equation}
\hat{L}_{t}=e^{it\hat{g}}\label{eq:exp-Lt}
\end{equation}
with the $g$'s being hermitian \citep{Esposito}. It turns out that
if $\{L_{t}\}$ is a \emph{unitary} group, weak and strong continuities
are equivalent \citep{deOliviera}, and thus if the one-parameter group
of \emph{unitary} operators $\{L_{t}\}$ is weakly continuous, one
can write its elements as exponentiation of some hermitian operators
(generators $g$ of $\{L_{t}\}$) on $\mathscr{H}$ as in (\ref{eq:exp-Lt}),
or similarly as in (\ref{eq:exp-q-p-qt}). Thus, while the Weyl relation
(\ref{eq:Weyl-main}) is always valid, only for one-parameter group
of unitary operators (whose generators are well-defined on $\mathscr{H}$)
one can recover the algebra of the generators similar to (\ref{eq:qt-commut}).
Also note that, since $g$'s are the generators of the \emph{infinitesimal}
transformations (while the group members $L_{t}$ generate finite
transformations), the existence (or lack) of generators on $\mathscr{H}$
means the existence (or lack) of infinitesimal transformations. As
a sidenote we mention that a one-parameter unitary group is bounded
if it is weakly continuous, hence the connection between boundedness
and continuity.

The Schr\"{o}dinger representation is a representation of the Weyl
relation (\ref{eq:Weyl-main}) which is weakly continuous. The celebrated
Stone-von Neumann theorem \citep{Stone,vonNeumann,Hall-2013} states
that every irreducible representation of the Weyl relations in which
the operators are unitary and weakly continuous, are unitarily equivalent
to each other and to the Schr\"{o}dinger representation.

\subsection{Polymer representation: kinematics}

In cases where the above condition of weak continuity is not valid anymore, the resulting
representations of the Weyl relation is not unitarily equivalent to
the Schr\"{o}dinger representation \citep{Hall-2013}. The polymer
representation is such a representation for which (some of) the generators
$g$ are not well-defined on $\mathscr{H}$. This means that the theory
based on this representation does not admit these infinitesimal transformations
and only contains the finite ones. The full Hilbert space, $\mathcal{H}_{\textrm{poly}}$, then, possesses an uncountable
orthonormal basis such that 
\begin{equation}
\langle\alpha|\beta\rangle=\delta_{\alpha,\beta},\,\,\,\,\,\,\,\,\,\,\alpha,\beta\in\mathbb{R},\label{eq:in-prod-H-poly}
\end{equation}
where $\delta_{\alpha,\beta}$ is the Kronecker delta.

As we mentioned before, while one can classically write $U=e^{\frac{i}{\hbar}\mu q}$
and $V=e^{\frac{i}{\hbar}\lambda p}$, this is not always allowed
quantum mechanically for both of these operators. Thus, in polymer
representation, one usually chooses one of the two ``polarizations'':
the $q$-polarization in which $\hat{q}$ (but not $\hat{p}$) is
well-defined and we have 
\begin{align}
\hat{U}_{\mu}|q\rangle= & e^{\frac{i}{\hbar}\mu q}|q\rangle,\label{eq:q-pol-1}\\
\hat{V}_{\lambda}|q\rangle= & |q-\lambda\rangle,\label{eq:q-pol-2}
\end{align}
and the $p$-polarization in which $\hat{p}$ (but not $\hat{q}$)
is well-defined where 
\begin{align}
\hat{U}_{\mu}|p\rangle= & |p-\mu\rangle,\label{eq:p-pol-1}\\
\hat{V}_{\lambda}|p\rangle= & e^{\frac{i}{\hbar}\lambda p}|p\rangle.\label{eq:p-pol-2}
\end{align}
One can see that in the $q$-polarization, $V_{\lambda}$ is not weakly
continuous, $\lim_{\lambda\rightarrow0}\langle q|\hat{V}_{\lambda}|q\rangle\neq\langle q|\hat{V}_{\lambda=0}|q\rangle$,
since 
\begin{equation}
\lim_{\lambda\rightarrow0}\langle q|\hat{V}_{\lambda}|q\rangle=\lim_{\lambda\rightarrow0}\langle q|q-\lambda\rangle=0,
\end{equation}
while 
\begin{equation}
\langle\mu|\hat{V}_{\lambda=0}|\mu\rangle=\Vert\hat{V}_{0}\Vert=1,
\end{equation}
where the inner products have been taken using \eqref{eq:in-prod-H-poly}. This proves our claim that while the infinitesimal generator $\hat{q}$
exists and thus quantum mechanically we can write $\hat{U}_{\mu}=e^{\frac{i}{\hbar}\mu\hat{q}}$,
the infinitesimal generator $\hat{p}$ does not exist and thus we
only have finite translations (with steps $\lambda$) in $q$ space
generated by $V_{\lambda}$ as can be seen from (\ref{eq:q-pol-2}).
More precisely, once one fixes a certain $\lambda$ as a free parameter
of the theory, and starts from an initial state $|q_{0}\rangle$,
it is seen from (\ref{eq:q-pol-2}) (and also (\ref{eq:q-pol-1}))
that the wave functions $\langle q|\Psi\rangle$ are restricted to
the lattice points $\{q_{n}|q_{n}=q_{0}+n\lambda,\,n\in\mathbb{Z}\}$
and the eigenvalues of the operator $\hat{q}$ are discrete. On the
contrary, values of $p$ corresponding to the basis $|p\rangle$ are
not discrete but take values on a circle, i.e., $-\frac{\pi\hbar}{\lambda}\leq p<\frac{\pi\hbar}{\lambda}$
(see for example appendix A of \citep{Hugo-Saeed-Daniel}). The discreteness
in $q$ is thus inherent in polymer type of theories. As a result
of the above discussion, in this polarization it is usual to write
the $|q\rangle$ basis as $|q_{n}\rangle$ which are the members of
a countable basis of the corresponding Hilbert space $\mathcal{H}_{q_{0}}$.
Note, however, that the full polymer Hilbert space is 
\begin{equation}
\mathcal{H}_{\textrm{poly}}=\bigoplus_{0\leq q_{0}<\lambda}\mathcal{H}_{q_{0}}.
\end{equation}
Thus $\mathcal{H}_{q_{0}}$ is a separable super-selected sector of
$\mathcal{H}_{\textrm{poly}}$ while, as mentioned above, the full non-separable polymer
Hilbert space, possesses an inner product as in (\ref{eq:in-prod-H-poly})

For the $p$-polarization, (\ref{eq:p-pol-1}) and (\ref{eq:p-pol-2}),
things are reversed but the ideas are essentially the same. There,
$\hat{U}_{\mu}$ is not weakly continuous and thus its generator $\hat{q}$
is not well-defined on the Hilbert space. For a fixed $\mu$ we have
a lattice in $p$ space such that starting from an initial state $|p_{0}\rangle$,
the wave functions $\langle p|\Psi\rangle$ are restricted to the
lattice points $\{p_{m}|p_{m}=p_{0}+m\mu,\,m\in\mathbb{Z}\}$. Also
$q$ takes continuous values on a circle, $-\frac{\pi\hbar}{\mu}\leq q<\frac{\pi\hbar}{\mu}$,
and the corresponding Hilbert space $\mathcal{H}_{p_{0}}$ is a super-selected
sector of $\mathcal{H}_{\textrm{poly}}$, such that $\mathcal{H}_{\textrm{poly}}=\bigoplus_{0\leq p_{0}<\mu}\mathcal{H}_{p_{0}}$.

\subsection{Polymer representation: dynamics}

Let us consider the $q$-polarization \citep{Hugo-Saeed-Daniel,Hugo-Saeed-Daniel-MG14} where $\hat{U}_{\mu}=e^{\frac{i}{\hbar}\mu\hat{q}}$
and  
\begin{align}
\hat{q}|q_{n}\rangle= & q_{n}|q_{n}\rangle,\label{eq:q-descrt-polarz-1}\\
\hat{V}_{\lambda}|q_{n}\rangle= & |q_{n}-\lambda\rangle,\label{eq:q-descrt-polarz-2}\\
\hat{q}|p\rangle= & \frac{\hbar}{i}\partial_{p}|p\rangle,\\
\hat{V}_{\lambda}|p\rangle= & e^{\frac{i}{\hbar}\lambda p}|p\rangle.\label{eq:q-descrt-polarz-3}
\end{align}
Since in this case the generator $\hat{p}$ does not exist, we need
to construct its analog to be able to represent e.g. the kinetic term
$p^{2}/2m$ in the Hamiltonian. It turns out that it is better to
start from an analog of $\hat{p}^{2}$. Classically, we have the following
approximation 
\begin{equation}
e^{\frac{i\lambda p}{\hbar}}+e^{-\frac{i\lambda p}{\hbar}}\approx2-\frac{\lambda^{2}p^{2}}{\hbar^{2}},\,\,\,\,\,\,\,\,\,p\ll\frac{\hbar}{\lambda},
\end{equation}
which together with (\ref{eq:q-descrt-polarz-3}) can be used to give
the analog of $\hat{p}^{2}$ as 
\begin{equation}
\widehat{p_{\lambda}^{2}}=\frac{\hbar^{2}}{\lambda^{2}}\left(2-\hat{V}_{\lambda}-\hat{V}_{-\lambda}\right).\label{eq:P2-poly}
\end{equation}
Its action on $|p\rangle$ basis can then be computed to yield 
\begin{equation}
\widehat{p_{\lambda}^{2}}|p\rangle=\frac{4\hbar^{2}}{\lambda^{2}}\left(\sin^{2}\left(\frac{\lambda p}{2\hbar}\right)\right)|p\rangle.
\end{equation}
Using this as a guide, we can define the analog of $\hat{p}$ such
that 
\begin{equation}
\widehat{p_{\lambda}}|p\rangle=\frac{\hbar}{\lambda}\left(\sin\left(\frac{\lambda p}{\hbar}\right)\right)|p\rangle.
\end{equation}
This can be achieved by defining the analog of $\hat{p}$ as 
\begin{equation}
\widehat{p_{\lambda}}=\frac{\hbar}{2i\lambda}\left(\hat{V}_{\lambda}-\hat{V}_{-\lambda}\right).
\end{equation}
The same construction can be used to represent the analog of $\hat{q}$
and $\hat{q}^{2}$ in the $p$-polarization where $\hat{q}$ is not
well-defined on the Hilbert space.

As a result of these definitions, the Hamiltonian operator of a free
particle can be represented as
\begin{equation}
\hat{H}=\frac{1}{2m}\frac{\hbar^{2}}{\lambda^{2}}\left(2-\hat{V}_{\lambda}-\hat{V}_{-\lambda}\right)
\end{equation}
whose action on the $|p\rangle$ basis yields 
\begin{equation}
\hat{H}|p\rangle=\frac{2\hbar^{2}}{m\lambda^{2}}\left(\sin^{2}\left(\frac{\lambda p}{2\hbar}\right)\right)|p\rangle.
\end{equation}

\section{Polymer deparametrized propagator\label{sec:poly-depar-prop}}

\subsection{Deparametrization}

There are certain physical systems whose action is invariant under
reparametrization, $\tau\rightarrow f(\tau)$, of the time variable $\tau$, and thus the physics of the model will not change. So the mentioned transformation
is a gauge transformation and the parameter $\tau$ is a gauge parameter
which does not represent the true physical time. Thus the evolution
with respect to $\tau$ is just unfolding gauge transformations. Expectedly,
these systems are totally constrained which means that the Hamiltonian
can be written as
\begin{equation}
H=\int d^{3}xN^{i}\mathcal{C}_{i},
\end{equation}
where $N^{i}$ are Lagrange multipliers and $\mathcal{C}_{i}$ are (first class)
constraints of the system. In order to get a true physical evolution
(i.e. relational evolution), one should fully fix the gauge in these systems.
Then it turns out that at least one of the introduced gauge fixing
conditions $\chi_{j}=0$, should explicitly depend on $\tau$ and
hence $\tau$ can be expressed as a combination of canonical variables
$\tau=t(Q,P)$, where $Q,P$ symbolically stands for a full set of phase space variables which has a dimension equal to or greater than four. Such a time variable $t$ is called an internal
time. 

In some of these models, one can explicitly find (a combination of)
canonical variable(s) as above, which has a monotonic gauge-dependent
relationship with $\tau$. This (combination of) canonical variable(s)
can then be taken as the internal time\footnote{In principle one can have several internal times $T^{I}$. Here we
assume there is only one.}, and using a canonical transformation $(Q,P)\rightarrow(t,\pi,q,p)$,
such that $t(Q,P)\rightarrow t$ and $\pi$ is conjugate to $t$,
the system can be ``deparametrized'' and written in the following
form\footnote{There are cases in which $h$ depends also on the physical time $T$, but models studied here do not have this property.}
\begin{equation}
\mathcal{C}=G(\pi)-h(q,p).\label{eq:generic-deparametr}
\end{equation}
This deparametrization can be done in two ways, either $G(\pi)=\pi$
or $G(\pi)=\pi^{2}$ (or some linear, or quadratic polynomials of
$\pi$ respectively). The former case is called non-relativistic while
the latter one is called relativistic deparametrization. Note that these terms actually
refer to the type of deparametrization not the physical model itself.
They are called like this due to the similarity of their form to relativistic
and non-relativistic dispersion relation (considering $\pi$ as the
``energy'' which is the momentum of ``time''). 

Now one has two choices: either use the non-fixed theory and find
the gauge time evolution with respect to $\tau$ using constraint
$\mathcal{C}$ for a function $f(q,p,t,\pi)$
\begin{align}
\frac{\partial f(q,p,t,\pi)}{\partial\tau}=\dot{f}(q,p,t,\pi)= & \left\{ f(q,p,t,\pi),\mathcal{C}\right\} \nonumber \\
= & \left\{ f(q,p,t,\pi),G(\pi)\right\} -\left\{ f(q,p,t,\pi),h(q,p)\right\} \nonumber \\
= & \frac{\partial f}{\partial t}\frac{\partial G}{\partial\pi}-\{f,h\},
\end{align}
or go to the reduced phase space where $\frac{\partial f(q,p,t,\pi)}{\partial\tau}=0$
and use the full gauge fixing condition, i.e. solving $\mathcal{C}=0$
and get (using above and the form of $\pi(q,p)$ from solving $\mathcal{C}=0$)
\begin{equation}
\frac{\partial f}{\partial t}=\frac{1}{\frac{\partial G}{\partial\pi}}\{f,h\}=\begin{cases}
\{f,h\} & \textrm{non-relativistic case}\\
\frac{1}{2\pi}\{f,h\}=\frac{1}{2\sqrt{h}}\{f,h\}=\left\{ f,\sqrt{h}\right\}  & \textrm{relativistic case}
\end{cases}
\end{equation}
We see that in the gauge fixed method, in both case above, the evolution
of $f$ with respect to $t$ is given by $\pi$ (after solving it
in terms of $h$ from $\mathcal{C}=0$). Hence $\pi(q,p)=h$ or $\pi(q,p)=\sqrt{h}$
provide us with the true evolution. We thus conclude that, from (\ref{eq:generic-deparametr}),
for a non-relativistic deparametrized constraint we have
\begin{equation}
\mathcal{C}=\pi-h(q,p),\label{eq:non-rel-deparametr}
\end{equation}
with $\pi=h(q,p)$ being the true Hamiltonian, while for a relativistic
one we get
\begin{equation}
\mathcal{C}=\pi^{2}-h(q,p),\label{eq:rel-deparametr}
\end{equation}
where the true Hamiltonian is $\pi=\sqrt{h(q,p)}$. Later we will
see that these actually correspond to the form of deparametrized constraints
of non-relativistic and relativistic particles\footnote{In these cases the notion of relativistic and non-relativistic deparametrizations
coincide with the properties of the physical systems themselves.}.

On the other hand, if one has a system with a true Hamiltonian $\pi(q,p)$
that generates true evolution with respect to (an internal) time $t$,
under certain conditions it is possible to extend the phase space
to also include $t$ and its conjugate $\pi$, and then turn the system
into a gauge system by deriving a constraint in deparametrized from
which gives the evolution of the system with respect to a non-physical
time parameter $\tau$ which is of course a gauge transformation.

\subsection{Non-relativistic and relativistic particles as constrained systems\label{subsec:Non-rel-rel-constraints}}

Following what we mentioned in the last subsection, we can write both
non-relativistic and relativistic particles as constrained system.
First consider the non-relativistic case. The action is simply
\begin{equation}
S=\int dt\frac{1}{2}m\left(\frac{dx}{dt}\right)^{2}.
\end{equation}
This is a system with true Hamiltonian
\begin{equation}
H=\frac{p_{x}^{2}}{2m}
\end{equation}
which generates true physical evolution with respect to $t$. In order
to enlarge the phase space by including $t$ in it, we introduce the
non-physical variable $\tau$ such that $\tau=f(t)$, to get
\begin{equation}
S=\int d\tau\frac{1}{2}m\frac{\left(\frac{dx}{d\tau}\right)^{2}}{\frac{dt}{d\tau}}=\int d\tau\frac{1}{2}m\frac{\dot{x}^{2}}{\dot{t}},
\end{equation}
where now $t$ is a canonical variable and (gauge) evolution is with
respect to $\tau$. Canonical analysis of the theory shows that we
have
\begin{align}
p_{x}= & \frac{\partial L}{\partial\dot{x}}=\frac{m\dot{x}}{\dot{t}}\\
p_{t}= & \frac{\partial L}{\partial\dot{t}}=-\frac{1}{2}\frac{m\dot{x}^{2}}{\dot{t}^{2}},
\end{align}
and the Hamiltonian is
\begin{equation}
H=p_{x}\dot{x}+p_{t}\dot{t}-L=\frac{m\dot{x}^{2}}{\dot{t}}-\frac{1}{2}\frac{m\dot{x}^{2}}{\dot{t}}-\frac{1}{2}m\frac{\dot{x}^{2}}{\dot{t}}=0,
\end{equation}
which shows that we have a totally constrained system. The system
exhibits a constraint that from definitions of $p_{x}$ and $p_{t}$
above reads
\begin{equation}
\mathcal{C}=p_{t}+\frac{p_{x}^{2}}{2m}=0.\label{eq:non-rel-constraint}
\end{equation}
Now we can see that this has the non-relativistic form of a deparametrized
constraint (\ref{eq:non-rel-deparametr}) with $\pi=p_{t}$ and thus
the true Hamiltonian after gauge fixing is $\pi=p_{t}=-\frac{p_{x}^{2}}{2m}$
which generates true evolution (with a minus sign) with respect to
$t$ as expected.

In the case of the relativistic particle, we have
\begin{equation}
S=\int dt\ m\sqrt{-\eta_{ab}\frac{dx^{a}}{dt}\frac{dx^{b}}{dt}}.
\end{equation}
Here the system is already constrained and is reparametrization-invariant
as can be seen by the fact that introducing $t=f(s)$ yields the same
action
\begin{equation}
S=\int ds\,m\sqrt{-\eta_{ab}\frac{dx^{a}}{ds}\frac{dx^{b}}{ds}}.
\end{equation}
Counting $t$ as a canonical variables and writing $\dot{x}^{a}=\frac{dx^{a}}{ds}$,
one can compute the momenta 
\begin{equation}
p_{a}=\frac{\partial L}{\partial\dot{x}^{a}}=-m\frac{\eta_{ab}\dot{x}^{b}}{\sqrt{-\eta_{ab}\frac{dx^{a}}{ds}\frac{dx^{b}}{ds}}}
\end{equation}
and it is easy to see that
\begin{equation}
H=p_{a}\dot{x}^{a}-L=0,
\end{equation}
hence a constraint system. The constraint is 
\begin{equation}
\mathcal{C}=\eta^{ab}p_{a}p_{b}+m^{2}=0,
\end{equation}
which is the famous relativistic dispersion relation. This can be
written as
\begin{equation}
\mathcal{C}=p_{t}^{2}-p_{x}^{2}-m^{2}=0\label{eq:rel-constraint}
\end{equation}
which not surprisingly takes the form (\ref{eq:rel-deparametr}) with
$G(\pi)=\pi^{2}=p_{t}^{2}$ and the true Hamiltonian in this case
(after complete gauge fixing) is $\pi=p_{t}=\sqrt{p_{x}^{2}+m^{2}}$.
Later we will use (\ref{eq:non-rel-constraint}) and (\ref{eq:rel-constraint})
to study each case.

\subsection{Quantization and group averaging}

To quantize a classical theory a la Dirac, one represents the classical
canonical variables and their algebra on a vector space such that
there are no anomalies, i.e. the quantum algebra $[\hat{q},\hat{p}]$
mimics the classical algebra $\{q,p\}$, and then equip this vector
space with an inner product and Cauchy-complete it to get a Hilbert
space. Given that the classical phase space functions can now be represented
as operators on this Hilbert space, one can proceed to compute the
desired quantities using these states, operators and the inner product.

If the classical theory has first class constraints $\mathcal{C}_{I}$,
i.e. it is a gauge theory, then the resulting Hilbert space is actually
not the Hilbert space $\mathscr{H}_{\textrm{phys}}$ of physical states
$|\psi_{\textrm{phys}}\rangle$, but in general it is a bigger space
called the kinematical Hilbert space $\mathscr{H}_{\textrm{kin}}$
with states $|\psi_{\textrm{kin}}\rangle$, not all of them corresponding
to the physical solutions. To be able to derive physical results, one
has to find $\mathscr{H}_{\textrm{phys}}$. This means finding a subset
of states of $\mathscr{H}_{\textrm{kin}}$\footnote{In general, the set $\left\{ |\psi\rangle_{\textrm{phys}}\right\} $
are not needed and can not be in $\mathscr{H}_{\textrm{kin}}$, hence
the word ``subset'' here may be misleading. This is specially
the case when the gauge group is non-compact. However, in that case,
$\left\{ |\psi\rangle_{\textrm{phys}}\right\} $ can exist in a certain
algebraic dual of a dense subset of $\mathscr{H}_{\textrm{kin}}$.
See for example \citep{Thiemann} for details.} that are physical, and then define an inner product between them
(physical inner product $\langle\cdot|\cdot\rangle_{\textrm{phys}}$)
so that this subspace of states becomes a Hilbert space. 

The set $\left\{ |\psi_{\textrm{phys}}\rangle\right\} $ are the ones
that are annihilated under the action of the quantum version of the
constraints $\hat{\mathcal{C}}_{I}$. The reason for this criterion
is that since the constraints generate infinitesimal gauge (i.e. non-physical)
transformations, physical states should remain invariant under these
transformations \citep{Claudio}. More precisely, since classical
first class constraints commute with each other weakly (i.e. on the
constraint surface)
\begin{equation}
\{\mathcal{C}_{I},\mathcal{C}_{J}\}\approx0\Rightarrow\{\mathcal{C}_{I},\mathcal{C}_{J}\}=\xi_{IJ}^{K}\mathcal{C}_{K},
\end{equation}
with $\xi_{IJ}^{K}$ being structure constants, and since the algebra
is represented in an anomaly-free manner, then one has 
\begin{equation}
\left[\hat{\mathcal{C}}_{I},\hat{\mathcal{C}}_{J}\right]|\psi_{\textrm{phys}}\rangle=0\Rightarrow\hat{\mathcal{C}}_{K}|\psi_{\textrm{phys}}\rangle=0.\label{eq:Dirac-act-const-phys}
\end{equation}
This way one can also see that the physical states $|\psi_{\textrm{phys}}\rangle$
are invariant under the action of the associated gauge group elements
$\hat{U}=e^{-i\alpha^{I}\hat{\mathcal{C}}_{I}}$
\begin{equation}
\hat{U}|\psi_{\textrm{phys}}\rangle=|\psi_{\textrm{phys}}\rangle
\end{equation}
as desired, with $\alpha^{I}$ the Lagrange multipliers that are the
parameters of the group. If these states are found, then one should
proceed to define an inner product between them to construct $\mathscr{H}_{\textrm{phys}}$.

Solving quantum constraints however, is not always an easy task, and
when one can not solve (\ref{eq:Dirac-act-const-phys}) for the states
$|\psi_{\textrm{phys}}\rangle$, one obviously cannot define an inner
product between them. So other methods are to be pursued. One way
to deal with this issue is the so-called group averaging technique
\citep{Marolf-RAQ-1,Marolf-RAQ-2,Thiemann}. Not only this method
solves for physical states in some sense, it also provides us with
a physical inner product given that the kinematical space already
has one, i.e. given that we already have a $\mathscr{H}_{\textrm{kin}}$
with $\langle\cdot|\cdot\rangle_{\textrm{kin}}$. We do not go into
the details but just mention a few points that are important to us
in this paper.

Given that $\hat{U}(\alpha)=e^{-i\alpha\hat{\mathcal{C}}}$ are the
members of the gauge group, one writes physical states as
\begin{equation}
\langle\psi_{\textrm{phys}}|=\int d\alpha\langle\psi_{\textrm{kin}}|\hat{U}^{\dagger}(\alpha)\label{eq:grp-avrg}
\end{equation}
and then, using the kinematical inner product, the physical inner
product is defined as
\begin{equation}
\langle\psi_{\textrm{phys}}|\phi_{\textrm{phys}}\rangle_{\textrm{phys}}:=\int d\alpha\langle\psi_{\textrm{kin}}|\hat{U}^{\dagger}(\alpha)|\phi_{\textrm{kin}}\rangle_{\textrm{kin}}.\label{eq:inn-prod-general}
\end{equation}
This inner product has certain properties which amounts to solving
for the states for which $\hat{U}(\alpha)|\psi\rangle=|\psi\rangle$.

Finally note that in the non-relativistic deparametrization case,
(\ref{eq:non-rel-deparametr}) or (\ref{eq:non-rel-constraint}),
if we represent the ``time momentum'' as a derivative $\hat{\pi}=i\hbar\frac{\partial}{\partial t}$
and use the condition of physical states (\ref{eq:Dirac-act-const-phys}),
we will get
\begin{equation}
\hat{\mathcal{C}}|\psi_{\textrm{phys}}\rangle=0\Rightarrow i\frac{\partial}{\partial t}|\psi_{\textrm{phys}}\rangle=\hat{h}|\psi_{\textrm{phys}}\rangle
\end{equation}
which is just the Schr\"{o}dinger's equation given that $\hat{h}$ is
the true Hamiltonian operator generating physical evolution with respect
to the physical time $t$. In the case of relativistic deparametrization
(\ref{eq:rel-deparametr}) or (\ref{eq:rel-constraint}), by the same
argument and representation we get
\begin{equation}
\hat{\mathcal{C}}|\psi_{\textrm{phys}}\rangle=0\Rightarrow-\hbar^{2}\frac{\partial^{2}}{\partial t^{2}}|\psi_{\textrm{phys}}\rangle=\hat{h}|\psi_{\textrm{phys}}\rangle,
\end{equation}
which for the case of a particle gives the equivalent of the Kelin-Gordon
equation. So in both cases, one can regain a physical evolution in
terms of a true Hamiltonian operator that generates time evolution
with respect to an internal physical time $t$.

\subsection{Polymer path integral}
Most of the interesting properties of quantum systems are expressed
in terms of transition amplitudes 
\begin{equation}
\langle\psi_{f}|\hat{U}(t_{f},t_{i})|\psi_{i}\rangle,\,\,\,\,\,\,\,\,\,\,\,\,\,\,\,\,\,t_{i}<t_{f},
\end{equation}
between initial and final states $|\psi_{i}\rangle$ and $|\psi_{f}\rangle$
respectively, with an evolution operator $\hat{U}\left(t_{f},t_{i}\right)$.
Using the completeness relations of some basis, say $|q\rangle$, one
can write
\begin{equation}
\int dq_{f}\int dq_{i}\langle\psi_{f}|q_{f}\rangle\langle q_{f}|\hat{U}(t_{f},t_{i})|q_{i}\rangle\langle q_{i}|\psi_{i}\rangle=\int dq_{f}\int dq_{i}\psi_{f}^{*}(q_{f})\psi_{i}(q_{i})K\left(q_{f},t_{f};q_{i},t_{i}\right),
\end{equation}
where we have defined the kernel $K\left(q_{f},t_{f};q_{i},t_{i}\right)$
as the matrix elements of the evolution operator 
\begin{equation}
K\left(q_{f},t_{f};q_{i},t_{i}\right)=\langle q_{f}|\hat{U}(t_{f},t_{i})|q_{i}\rangle.\label{eq:kernel-matrix-U}
\end{equation}
The kernel itself may be written as a path integral
\begin{equation}
K\left(q_{f},t_{f};q_{i},t_{i}\right)=\int\mathscr{D}'q\mathscr{D}'pe^{\frac{i}{\hbar}S'}
\end{equation}
as we  will see in details in what follows for different cases. Note that In this case, $S'$ and $\mathcal{D}'q\mathcal{D}'p$ are not the classical action and the usual measure that appears in the Schr\"{o}dinger quantization, hence the primes used for them.

Given that the Hamiltonian is time independent, the kernel in the
form (\ref{eq:kernel-matrix-U}) can be interpreted as the transition
amplitude
\begin{equation}
K\left(q_{f},t_{f};q_{i},t_{i}\right)=\langle q_{f},t_{f}|q_{i},t_{i}\rangle\label{eq:kernel-inner-q-t}
\end{equation}
between the kets $|q_{i},t_{i}\rangle=e^{\frac{i}{\hbar}\hat{H}t_{i}}|q_{i}\rangle$
and $|q_{f},t_{f}\rangle=e^{\frac{i}{\hbar}\hat{H}t_{f}}|q_{f}\rangle$.
Note that these are not the Schr\"{o}dinger-picture time evolution
of states $|q_{i}\rangle$ and $|q_{f}\rangle$, since they have the
wrong sign in the exponent in front of them, but they are actually
eigenstates of the $\hat{q}$ operator in the Heisenberg picture at
corresponding times $t_{i}$ and $t_{f}$.

If the states $|q_{i},t_{i}\rangle$ and $|q_{f},t_{f}\rangle$ belong
to the $\mathscr{H}_{\textrm{phys}}$ of a single particle system,
a gauge system with only a single first class constraint
$\mathcal{C}$, the kernel due to (\ref{eq:inn-prod-general}) and
(\ref{eq:kernel-inner-q-t}) can be written as
\begin{equation}
K\left(q_{f},t_{f};q_{i},t_{i}\right)_{\textrm{phys}}=\langle q_{f},t_{f}|q_{i},t_{i}\rangle_{\textrm{phys}}=\int d\alpha\langle q_{f},t_{f}{}_{\textrm{kin}}|\hat{U}^{\dagger}(\alpha)|q_{i},t_{i}{}_{\textrm{kin}}\rangle_{\textrm{kin}}=\int_{\mathbb{R}}d\alpha\langle q_{f},t_{f}{}_{\textrm{kin}}|e^{i\alpha\hat{\mathcal{C}}}|q_{i},t_{i}{}_{\textrm{kin}}\rangle_{\textrm{kin}}.
\end{equation}
From now on we drop the ``phys'' and ``kin'' subscripts and they
are to be understood in the context. Using (\ref{eq:non-rel-deparametr})
and (\ref{eq:rel-deparametr}), the kernel for non-relativistic and
relativistic deparametrized cases can then be written as
\begin{equation}
K\left(q_{f},t_{f};q_{i},t_{i}\right)=\begin{cases}
\int_{\mathbb{R}}d\alpha\langle q_{f},t_{f}|e^{i\alpha\left(\hat{\pi}-\hat{h}\right)}|q_{i},t_{i}\rangle=\int_{\mathbb{R}}d\alpha\langle q_{f},t_{f}|e^{i\alpha\hat{\pi}}e^{-i\alpha\hat{h}}|q_{i},t_{i}\rangle, & \textrm{non-relativistic},\\
\int_{\mathbb{R}}d\alpha\langle q_{f},t_{f}|e^{i\alpha\left(\hat{\pi}^{2}-\hat{h}\right)}|q_{i},t_{i}\rangle=\int_{\mathbb{R}}d\alpha\langle q_{f},t_{f}|e^{i\alpha\hat{\pi}^{2}}e^{-i\alpha\hat{h}}|q_{i},t_{i}\rangle, & \textrm{relativistic},
\end{cases}
\end{equation}
since in general $\hat{\pi}$ and $\hat{h}$ commute. If we assume
that the states $|q,t\rangle$ can be written as a tensor product
$|q,t\rangle=|q\rangle\otimes|t\rangle$, which can be taken as a
consequence of $\hat{\pi}$ and $\hat{h}$ commuting, and choose a
polarization of the polymer representation such that the form of $\hat{h}$
(rather than $\hat{\pi}$) is changed from Schr\"{o}dinger to polymer
form (or equivalently only polymerize the pair $(q,p)$ and quantize
the pair $(\pi,t)$ using Schr\"{o}dinger representation), then we
can write
\begin{equation}
K_{\textrm{poly}}\left(q_{f},t_{f};q_{i},t_{i}\right)=\begin{cases}
\int_{\mathbb{R}}d\alpha\langle t_{f}|e^{i\alpha\hat{\pi}}|t_{i}\rangle\langle q_{f}|e^{-i\alpha\hat{h}}|q_{i}\rangle=\int_{\mathbb{R}}d\alpha A_{\pi}(t_{f},t_{i};\alpha)A_{\textrm{poly}}(q_{f},q_{i};\alpha), & \textrm{non-relativistic},\\
\int_{\mathbb{R}}d\alpha\langle t_{f}|e^{i\alpha\hat{\pi}^{2}}|t_{i}\rangle\langle q_{f}|e^{-i\alpha\hat{h}}|q_{i}\rangle=\int_{\mathbb{R}}d\alpha A_{\pi}(t_{f},t_{i};\alpha)A_{\textrm{poly}}(q_{f},q_{i};\alpha), & \textrm{relativistic},
\end{cases}\label{eq:Prop-Both}
\end{equation}
where $\langle f|\exp\left(\cdots\right)|i\rangle$ can be taken as
the probability amplitudes of transiting from $|i\rangle$ to $|f\rangle$
under the unitary transformation $\exp\left(\cdots\right)$. The amplitudes
have been factorized into two terms due to $[\hat{h},\hat{\pi}]=0$
and the resulting factorization $|q,t\rangle=|q\rangle\otimes|t\rangle$.
We have defined the polymer amplitude in each case as
\begin{equation}
A_{\textrm{poly}}(q_{f},q_{i};\alpha):=\langle q_{f}|e^{-i\alpha\hat{h}}|q_{i}\rangle.\label{eq:A-poly-def}
\end{equation}
The other amplitude
\begin{equation}
A_{\pi}(t_{f},t_{i};\alpha)=\begin{cases}
\langle t_{f}|e^{i\alpha\hat{\pi}}|t_{i}\rangle, & \textrm{non-relativistic}\\
\langle t_{f}|e^{i\alpha\hat{\pi}^{2}}|t_{i}\rangle, & \textrm{relativistic}
\end{cases}\label{eq:A-pi-Both}
\end{equation}
corresponds to the momentum of the physical time. In the rest of the
work, we will derive these amplitudes and subsequently the whole kernel
for three cases of a non-relativistic free particle, non-relativistic
particle in a box, and the relativistic particle.

\section{Generic polymer amplitude \label{sec:generic-amp-poly}}

If $\hat{h}$ in $A_{\textrm{poly}}(q_{f},q_{i};\alpha)$ in (\ref{eq:A-poly-def})
commutes with itself, which is the case for both free relativistic
and non-relativistic particles (see (\ref{eq:rel-constraint}) and
(\ref{eq:non-rel-constraint})), one can write the polymer amplitude
as
\begin{equation}
A_{\textrm{poly}}(q_{f},q_{i};\alpha)=\left\langle q_{f}\left|\left(e^{-i\alpha\frac{\hat{h}}{N}}\right)^{N}\right|q_{i}\right\rangle =\left\langle q_{f}\left|\underbrace{e^{-i\alpha\epsilon\hat{h}}\cdots e^{-i\alpha\epsilon\hat{h}}}_{N\,\textrm{times}}\right|q_{i}\right\rangle 
\end{equation}
with $\epsilon=\frac{1}{N}$. This is done so that we can proceed
with the Feynman procedure of writing the amplitude as a sum over
histories. Consequently this decomposition means that we are taking
$\hat{h}$ as an evolution operator and are considering the whole
(gauge) temporal interval as $\Delta\tau=N\epsilon=1$. Note that,
we are looking at these systems as constrained ones, since the evolution
of the system is generated by the constraint $\mathcal{C}$, hence
the whole evolution is a gauge transformation. So $t$ is one of the
canonical variables and the evolution steps are counted in $\tau$
rather than internal time $t$.

If the $|q\rangle$ basis is discrete, e.g. when we are in $q$-polarization
of the polymer representation, we have the identity $\hat{\mathbb{I}}_{q}=\sum_{n}|q_{n}\rangle\langle q_{n}|$,
and inserting $N-1$ of them in between the exponentials in above
yields
\begin{equation}
A_{\textrm{poly}}(q_{f},q_{i};\alpha)=A_{\textrm{poly}}(\bar{q}_{N},\bar{q}_{0};\alpha)=\sum_{\bar{q}_{1},\cdots,\bar{q}_{N-1}}\left\langle \bar{q}_{N}\left|e^{-i\alpha\epsilon\hat{h}}\right|\bar{q}_{N-1}\right\rangle \left\langle \bar{q}_{N-1}\left|e^{-i\alpha\epsilon\hat{h}}\right|\bar{q}_{N-2}\right\rangle \cdots\left\langle \bar{q}_{1}\left|e^{-i\alpha\epsilon\hat{h}}\right|\bar{q}_{0}\right\rangle ,\label{eq:A-poly-several-exp}
\end{equation}
where $q_{f}=\bar{q}_{N}$ and $q_{i}=\bar{q}_{0}$ and we maintain
that it is possible to have $\bar{q}_{k}=\bar{q}_{k\pm1}$ for some
or all of $k$. If $\bar{q}_{k-1}\neq\bar{q}_{k}$ we say that a position
transition has happened. This way we can express $A_{\textrm{poly}}(\bar{q}_{N},\bar{q}_{0};\alpha)$
in terms of smaller amplitudes $U_{k,l}=\left\langle \bar{q}_{k}\left|e^{-i\alpha\epsilon\hat{h}}\right|\bar{q}_{l}\right\rangle $,
\begin{equation}
A_{\textrm{poly}}(\bar{q}_{N},\bar{q}_{0};\alpha)=\sum_{\bar{q}_{1},\cdots,\bar{q}_{N-1}}\prod_{k=1}^{N}U_{k,k-1}.\label{eq:A-Prod-B}
\end{equation}
The picture here is one of a particle going from an initial $\bar{q}_{0}$
to a final $\bar{q}_{N}$ in $N$ steps, each step e.g. from $\bar{q}_{k-1}$
to $\bar{q}_{k}$ taken under a finite evolution $U_{k,k-1}$. This
way we have introduced a discrete path or history $\sigma_{N}=\left(\bar{q}_{0},\cdots,\bar{q}_{N}\right)$
for the particle. Typically at this point, one derives the path integral
by taking the limit $N\rightarrow\infty$. However this is not rigorous
in our case since we are in $q$-polarization of the polymer representation
and this means that the discreteness of $q$ in e.g. (\ref{eq:A-poly-several-exp})
is intrinsic and not just due to arbitrary non-physical discretization
of the interval $[q_{i},q_{f}]$ often used in deriving path integral
in ordinary quantum field theory. Thus we need to find a rigorous
manner to do this for our case. To do this, we reformulate the model
in terms of position transitions instead of time steps as follows.

To implement the position transition framework, we first assume that
in these $N$ steps, $M$ position transitions $\bar{q}_{k-1}\rightarrow\bar{q}_{k}$
may happen and we call these distinct positions $q_{k}$ without a
bar. Thus in these $N$ steps, we may have $M+1$ positions $q_{k}$
corresponding to $M$ transitions. However, we only put a restriction
on the set $\{q_{k}\}$ such that although no two consecutive $q_{k}$'s
can be the same, two non-consecutive ones, i.e. with $q_{j}$ and $q_{l}$
when $j\neq l\pm1$, can. So we have provided a discrete history $\sigma_{N}^{M}$
for the particle's position in the form of an ordered sequence
\begin{align}
\sigma_{N}^{M}= & \left(\underbrace{\bar{q}_{N_{M+1}-1},\ldots,\bar{q}_{N_{M}};\overbrace{\bar{q}_{N_{M}-1},\ldots,\bar{q}_{N_{M-1}};\ldots\ldots;\overbrace{\bar{q}_{N_{2}-1},\ldots,\bar{q}_{N_{1}};\underbrace{\bar{q}_{N_{1}-1},\ldots,\bar{q}_{0}}_{N_{1}}}^{N_{2}}}}_{N_{M+1}=N}^{N_{M}}\right)\nonumber \\
= & \left(\underbrace{q_{M},\ldots,q_{M};q_{M-1},\ldots,q_{M-1};\ldots\ldots;\overbrace{q_{1},\ldots,q_{1};\underbrace{q_{0},\ldots,q_{0}}_{N_{1}}}^{N_{2}}}_{N_{M+1}=N}\right),\,\,\,\,\,\,\,\,M\leq N,\,N_{k}>N_{k-1},
\end{align}
of a particle that starts from $\bar{q}_{0}=q_{0}$, and in $N_{1}-1$
steps (for the period $\tau=\left(N_{1}-1\right)\epsilon$) still
remains at $\bar{q}_{N_{1}-1}=q_{0}$ without any position transition.
Then, at this time, it makes a transition from there to $\bar{q}_{N_{1}}=q_{1}\neq q_{0}$,
and so on. At the $r$'th transition, the particle goes from $\bar{q}_{N_{r}-1}=q_{r-1}$
to $\bar{q}_{N_{r}}=q_{r}$. Note that the total number of transitions
$M$ is smaller or equal to the total number of steps $N$, and all
the $\bar{q}$ between (and equal to) $\bar{q}_{N_{r}}$ to $\bar{q}_{N_{r+1}-1}$
are equal to $q_{r+1}$. Also note that $\bar{q}_{k}=k\epsilon$ is
the position after $k$ \emph{steps}, while $q_{r}=N_{r}\epsilon$
is the position after the $r$'th \emph{transition}. This history
can be expressed more concisely in terms of two ordered sequences 
\begin{equation}
\sigma_{N}^{M}=\left\{ \left(q_{N},\cdots,q_{0}\right),\left(N_{M},\cdots,N_{1}\right)\right\} ,\,\,\,\,\,\,\,\,\,\,\,\,\,\,\,\,\,\,q_{k-1}\neq q_{k},\,N_{k}>N_{k-1}.
\end{equation}
This way, the probability amplitude of a certain history $\sigma_{N}^{M}$
is written as
\begin{equation}
A_{\textrm{poly}}\left(\sigma_{N}^{M}\right)=U_{M,M}^{N-N_{M}-1}U_{M,M-1}\cdots U_{1,1}^{N_{2}-N_{1}-1}U_{1,0}U_{0,0}^{N_{1}-1},\label{eq:A-poly-step-0}
\end{equation}
where now
\begin{equation}
U_{k,l}=\left\langle q_{k}\left|e^{-i\alpha\epsilon\hat{h}}\right|q_{l}\right\rangle .
\end{equation}
Next, we need to implement the sum over all the amplitudes corresponding
to all possible histories. This is done in three steps:
\begin{description}
\item [{Step~1}] We consider all the possible transitions at any possible
time step, e.g. the first transition may happen at any time between
$\tau=0$ (particle at $q_{0}$ only at $\tau=0$ and then goes from
$q_{0}$ to $q_{1}$) to $\tau=\left(N_{2}-1\right)\epsilon$ (particle
staying at $q_{0}$ for up to the end of $\tau=\left(N_{2}-2\right)\epsilon$,
at that time goes from $q_{0}$ to $q_{1}$ and remains there until
$\tau=\left(N_{2}-1\right)\epsilon$, and then goes from $q_{1}$
to $q_{2}$ such that at $\tau=N_{2}\epsilon$ it is at $q_{2}$),
and so forth. Summing over all these types of amplitudes for all
possible minima to maxima of $N_{i}$'s, we get 
\begin{equation}
A_{\textrm{poly}}^{N}(q_{M},\ldots q_{0};\alpha)=\sum_{N_{M}=M}^{N-1}\sum_{N_{M-1}=M-1}^{N_{M}-1}\cdots\sum_{N_{1}=1}^{N_{2}-1}A_{\textrm{poly}}\left(\sigma_{N}^{M}\right),\,\,\,\,\,\,\,\,\,\,\,\,N=N_{M+1}.\label{eq:A-poly-step-1}
\end{equation}
\item [{Step~2}] Here we sum over all the paths with exactly $M$ transitions
but with all possible intermediate values of $q_{k}$'s, given fixed
initial and final positions $q_{0}$ and $q_{M}$, and such that $q_{k-1}\neq q_{k}$
as mentioned before,
\begin{equation}
A_{\textrm{poly}}^{N}(M;\alpha)=\sum_{\underset{q_{k-1}\neq q_{k}}{q_{M-1},\ldots q_{1}}}A_{\textrm{poly}}^{N}(q_{M},\ldots q_{0};\alpha).\label{eq:A-poly-step-2}
\end{equation}
\item [{Step~3}] Finally we sum over all possible number of transitions
$0\leq M\leq N$, and since this amplitude does not depend on $N$,
we can take the limit $N\rightarrow\infty$ to get the full polymer
amplitude
\begin{equation}
A_{\textrm{poly}}(q_{f},q_{i};\alpha)=\lim_{N\rightarrow\infty}\sum_{M=0}^{N}A_{\textrm{poly}}^{N}(M;\alpha).\label{eq:A-poly-step-3}
\end{equation}
\end{description}

\subsubsection*{Free particle}

Since we are going to work with free particles, further simplifications
of the above amplitude are possible. In that case, since the operator
$\hat{h}$ in $U_{k,l}=\left\langle q_{k}\left|e^{-i\alpha\epsilon\hat{h}}\right|q_{l}\right\rangle $
only contains the polymer kinetic term with no potential term present
to introduce $q$ dependence, we will have
\begin{equation}
U_{0,0}=U_{1,1}=\cdots=U_{M,M}.
\end{equation}
Using this, (\ref{eq:A-poly-step-0}) will become
\begin{align}
A_{\textrm{poly}}\left(\sigma_{N}^{M}\right)= & U_{M,M-1}\cdots U_{1,0}\left[U_{M,M}^{N-N_{M}-1}\cdots U_{M,M}^{N_{2}-N_{1}-1}U_{M,M}^{N_{1}-1}\right],\,\,\,\,\,\,\,\,N=N_{M+1}\nonumber \\
= & \left[\prod_{k=1}^{M}U_{k,k-1}\right]U_{M,M}^{N-M-1}.\label{eq:A-poly-step-0-free-1}
\end{align}
On the other hand $U_{M,M}^{N-M-1}$ for large $N$ and finite $M$
is 
\begin{equation}
U_{M,M}^{N-M-1}=e^{-i\alpha h_{M,M}}\left[1+\mathcal{O}\left(N^{-1}\right)\right],
\end{equation}
where $h_{MM}=\left\langle q_{M}\left|\hat{h}\right|q_{M}\right\rangle $
and $\epsilon=N^{-1}$. One also gets
\begin{align}
U_{k,k-1}= & \left\langle q_{k}\left|e^{-i\alpha\epsilon\hat{h}}\right|q_{l}\right\rangle \nonumber \\
= & \left\langle q_{k}\left|1-i\alpha\epsilon\hat{h}+\mathcal{O}\left(\epsilon^{2}\right)\right|q_{l}\right\rangle \nonumber \\
= & -i\alpha\epsilon h_{k,k-1}\left[1+\mathcal{O}\left(N^{-2}\right)\right].
\end{align}
Substituting these into (\ref{eq:A-poly-step-0-free-1}) yields
\begin{equation}
A_{\textrm{poly}}\left(\sigma_{N}^{M}\right)=\left[\prod_{k=1}^{M}-i\alpha\epsilon h_{k,k-1}\left[1+\mathcal{O}\left(N^{-2}\right)\right]\right]e^{-i\alpha h_{M,M}}\left[1+\mathcal{O}\left(N^{-1}\right)\right].
\end{equation}
Now using this in the first step (\ref{eq:A-poly-step-1}), one arrives
at 
\begin{equation}
A_{\textrm{poly}}^{N}(q_{M},\ldots q_{0};\alpha)=\left(-i\alpha\epsilon\right)^{M}\left[\prod_{k=1}^{M}h_{k,k-1}\right]e^{-i\alpha h_{M,M}}\sum_{N_{M}=M}^{N-1}\sum_{N_{M-1}=M-1}^{N_{M}-1}\cdots\sum_{N_{1}=1}^{N_{2}-1}\left[1+\mathcal{O}\left(N^{-1}\right)\right].
\end{equation}
By changing $N_{j}\rightarrow N_{j}+j,\,\,\forall j$, the sums in
above can be cast into the following more convenient form,
\begin{equation}
A_{\textrm{poly}}^{N}(q_{M},\ldots q_{0};\alpha)=\left(-i\alpha\epsilon\right)^{M}\left[\prod_{k=1}^{M}h_{k,k-1}\right]e^{-i\alpha h_{M,M}}\sum_{N_{M}=0}^{N-M}\sum_{N_{M-1}=0}^{N_{M}}\cdots\sum_{N_{1}=0}^{N_{2}}\left[1+\mathcal{O}\left(N^{-1}\right)\right].
\end{equation}
In the following we first take the limit $N\rightarrow\infty$ (partially
performing step 3 above) and then proceed to sum over all possible
$q$ and then over all possible $M$. Each time interval without transition
$\tau_{j}$ has a time length of $\tau_{j}=N_{j}\epsilon$. Using
this, taking the limit $N_{j}\rightarrow\infty$, and noting $\epsilon\sum_{N_{j}}\rightarrow\int d\tau_{j}$,
we can convert the sums above into integrals
\begin{equation}
A_{\textrm{poly}}(q_{M},\ldots q_{0};\alpha)=\left(-i\alpha\right)^{M}\left[\prod_{k=1}^{M}h_{k,k-1}\right]e^{-i\alpha h_{M,M}}\int_{0}^{1}d\tau_{M}\cdots\int_{0}^{\tau_{2}}d\tau_{1}1=\frac{1}{M!}\left(-i\alpha\right)^{M}\left[\prod_{k=1}^{M}h_{k,k-1}\right]e^{-i\alpha h_{M,M}},
\end{equation}
where in the limit of the last integral $\epsilon\sum_{N_{M}=0}^{N-M}\rightarrow\int_{0}^{1}d\tau_{M}$
we have used the fact that the total interval is of unit length, $\tau=N\epsilon=1$,
and $M$ is finite and we can neglect its contribution:
\begin{equation}
\left(N-M\right)\epsilon\approx N\epsilon=1.
\end{equation}
Now we perform step 2 mentioned above and sum over all possible intermediate
$q_{k}$'s for a fixed $M$. This leads to
\begin{equation}
A_{\textrm{poly}}(q_{f},q_{i};M;\alpha)=\frac{1}{M!}\left(-i\alpha\right)^{M}e^{-i\alpha h_{M,M}}\sum_{\underset{q_{k-1}\neq q_{k}}{q_{M-1},\ldots q_{1}}}\left[\prod_{k=1}^{M}h_{k,k-1}\right].\label{eq:Poly-amp-semifin}
\end{equation}
Finally step 3 is completed by summing over all possible number of
transitions $M$ which yields
\begin{equation}
A_{\textrm{poly}}(q_{f},q_{i};\alpha)=\sum_{M=0}^{\infty}\frac{1}{M!}\left(-i\alpha\right)^{M}e^{-i\alpha h_{M,M}}\sum_{\underset{q_{k-1}\neq q_{k}}{q_{M-1},\ldots q_{1}}}\left[\prod_{k=1}^{M}h_{k,k-1}\right].\label{eq:Poly-amp-Fin}
\end{equation}
Now we are ready to use this machinery to compute the full propagator
of three types of system in the following sections.

\section{Non-relativistic particle\label{sec:non-rel-partcl}}

\subsection{Free particle\label{subsec:Free-particle}}

\begin{figure}
\includegraphics[scale=0.45]{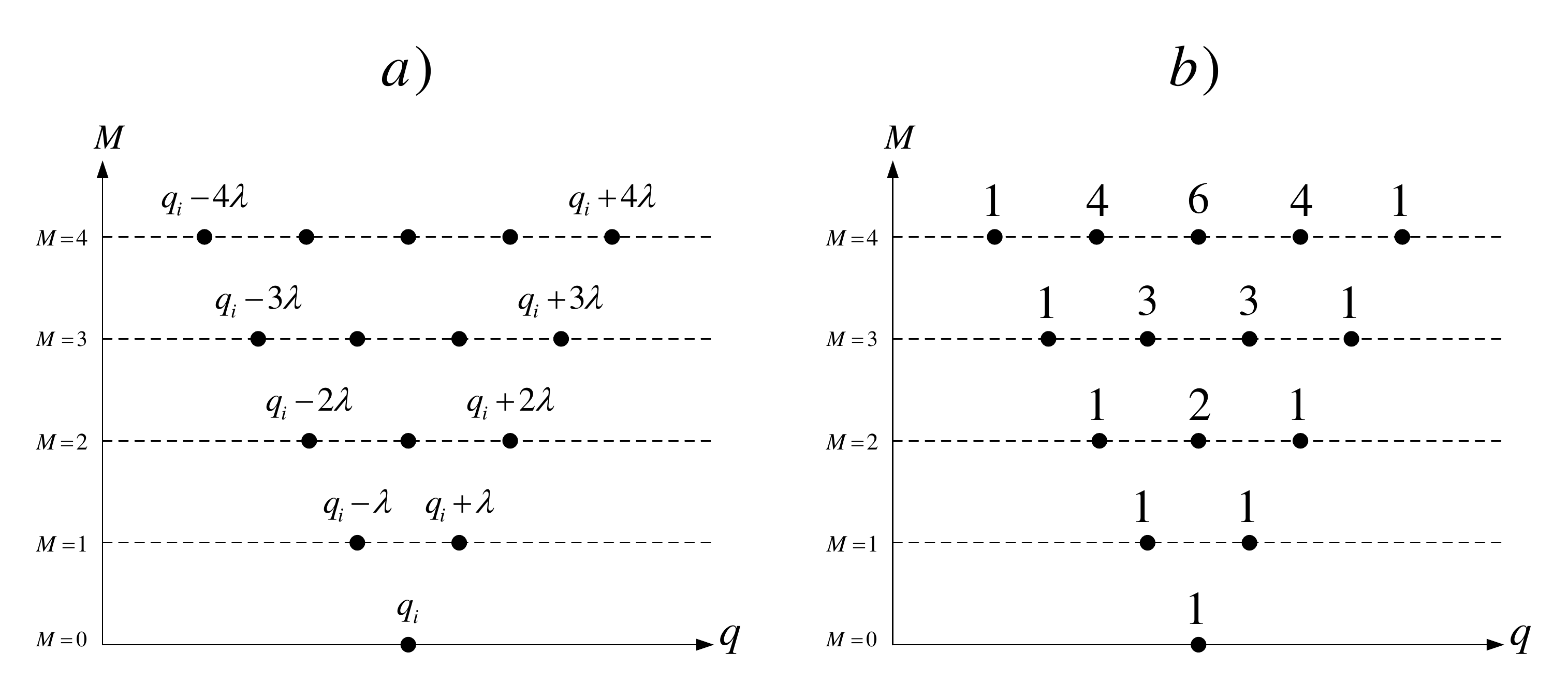}

\caption{a) The set of points for which the function $S(q_{f},q_{i};M)$ in
nontrivial. Only the coordinates of boundary points are written. b)
Nontrivial values of the function $S(q_{f},q_{i};M)$ that form the
values of a Pascal triangle.\label{fig:non-rel-triangle}}
\end{figure}

This system, written in deparametrized form, has two configuration
variables, $t$ and $q$. We will use the Schr\"{o}dinger representation
for $t$ and its canonical conjugate $p_{t}$, while $q$ and its
counterpart $V_{\lambda}$ are quantized using polymer representation.
Their algebras read
\begin{equation}
\left[\hat{t},\hat{p}_{t}\right]=i\hbar,\,\,\,\,\,\,\,\,\,\,\,\,\,\,\,\,\,\,\,\left[\hat{q},\hat{V}_{\lambda}\right]=-\frac{\lambda}{\hbar}\hat{V}_{\lambda},\,\,\,\,\,\,\,\,\,\,\,\,\,\,\,\,\,\,\,\left[\hat{t},\hat{q}\right]=0.\label{eq:Algebra-t,q}
\end{equation}
We derive the full polymer propagator of this model using (\ref{eq:non-rel-constraint})
and (\ref{eq:Prop-Both})-(\ref{eq:A-pi-Both}). First notice that
as mentioned before and as can be seen from (\ref{eq:non-rel-constraint})
and (\ref{eq:non-rel-deparametr}), the amplitude $A_{\pi}$ for this
system, using the identity $\hat{\mathbb{I}}_{p_{t}}=\int_{\mathbb{R}}dp_{t}|p_{t}\rangle\langle p_{t}|$,
is
\begin{equation}
\langle t_{f}|e^{i\alpha\hat{\pi}}|t_{i}\rangle=\langle t_{f}|e^{i\alpha\hat{p}_{t}}|t_{i}\rangle=\int_{\mathbb{R}}dp_{t}e^{i\alpha p_{t}}e^{\frac{i}{\hbar}p_{t}\left(t_{f}-t_{i}\right)}=\delta\left(\frac{t_{f}-t_{i}}{\hbar}+\alpha\right)\label{eq:non-rel-pt-part}
\end{equation}
Now we set to compute the polymer amplitude $A_{\textrm{poly}}(q_{f},q_{i};\alpha)$
using (\ref{eq:Poly-amp-Fin}). For this, we need to compute $h_{MM}$
and $h_{k,k-1}$. Looking at (\ref{eq:non-rel-constraint}) one can
see that in this case, $\hat{h}=-\widehat{\frac{p_{x}^{2}}{2m}}$.
Since we are in $q$-polarization, the generator corresponding to
$p_{x}$ does not exist on the Hilbert space and we should use (\ref{eq:P2-poly})
instead. This then leads to
\begin{equation}
\hat{h}=-\frac{1}{2m}\frac{\hbar^{2}}{\lambda^{2}}\left(2-\hat{V}_{\lambda}-\hat{V}_{-\lambda}\right).
\end{equation}
Using this and the form of the action of $\hat{V}_{\lambda}$, (\ref{eq:q-descrt-polarz-2}),
we can write
\begin{align}
h_{jk}= & -\frac{1}{2m}\frac{\hbar^{2}}{\lambda^{2}}\left\langle q_{j}\left|\left(2-\hat{V}_{\lambda}-\hat{V}_{-\lambda}\right)\right|q_{k}\right\rangle \nonumber \\
= & -\frac{1}{2m}\frac{\hbar^{2}}{\lambda^{2}}\left(2\delta_{q_{j},q_{k}}-\delta_{q_{j},q_{k}-\lambda}-\delta_{q_{j},q_{k}+\lambda}\right),
\end{align}
and hence 
\begin{equation}
h_{jk}=\begin{cases}
-\frac{\hbar^{2}}{m\lambda^{2}}, & j=k\\
\frac{1}{2m}\frac{\hbar^{2}}{\lambda^{2}}\left(\delta_{q_{j},q_{k}-\lambda}+\delta_{q_{j},q_{k}+\lambda}\right), & j\neq k.
\end{cases}
\end{equation}
Substituting these into (\ref{eq:Poly-amp-semifin}) yields
\begin{equation}
A_{\textrm{poly}}(q_{f},q_{i};M;\alpha)=\frac{1}{M!}e^{i\alpha\frac{\hbar^{2}}{m\lambda^{2}}}\left(-\frac{i\alpha}{2m}\frac{\hbar^{2}}{\lambda^{2}}\right)^{M}S(q_{f},q_{i};M),\label{eq:A-with-S}
\end{equation}
with
\begin{align}
S(q_{f},q_{i};M)= & \sum_{\underset{q_{k-1}\neq q_{k}}{q_{M-1},\ldots q_{1}}}\left[\prod_{k=1}^{M}\left(\delta_{q_{k},q_{k-1}-\lambda}+\delta_{q_{k},q_{k-1}+\lambda}\right)\right]\nonumber \\
= & \sum_{\underset{q_{k-1}\neq q_{k}}{q_{M-1},\ldots q_{1}}}\prod_{k=0}^{M-1}\left(\delta_{q_{M-k},q_{M-k-1}+\lambda}+\delta_{q_{M-k},q_{M-k-1}-\lambda}\right).
\end{align}
It can be shown by induction that this can be rewritten in a more
useful way
\begin{equation}
S(q_{f},q_{i};M)=\sum_{j=0}^{M}\begin{pmatrix}M\\
j
\end{pmatrix}\delta_{q_{M},q_{i}+\lambda\left(2j-M\right)},\label{eq:S-compact}
\end{equation}
where $q_{i}$ is the initial position of the particle. This function
has several important properties which will be used in the following
computations. The first property is invariance under translations
\begin{equation}
S(q_{2},q_{1};M)=S(q_{2}+r\lambda,q_{1}+r\lambda;M),\,\,\,\,\,\,\,\,\,\,\,r\in\mathbb{Z}.\label{eq:S-prop-1}
\end{equation}
Another one is the symmetry under change of the signs
\begin{equation}
S(q_{2},q_{1};M)=S(-q_{2},-q_{1};M).\label{eq:S-prop-2}
\end{equation}
The third property is what we call the ``Pascal triangle'' property
which is related to the points in which $S$ has
nontrivial values. More precisely, if the whole spatial interval is $q_{f}-q_{i}=\lambda L$,
then it is easily seen that $S$ in (\ref{eq:S-compact}) (and consequently
$A_{\textrm{poly}}(q_{f},q_{i};M;\alpha)$ in (\ref{eq:A-with-S}))
has only one nonvanishing term for which $j=\left(L+M\right)/2$.
Since $0\leq j\leq M$, this means that $|L|\leq M$. The graph of
spatial position of these terms for different $M$'s becomes like
a triangle, on each point of which, $S$ has a value corresponding
to the binomial coefficients, hence the Pascal triangle (Fig. (\ref{fig:non-rel-triangle})).

We now use these properties to simplify the polymer amplitude. First
we assume that $L$ is even. Then $A_{\textrm{poly}}(q_{f},q_{i};M;\alpha)$
will only have nontrivial values for an even $M=2k,\,\,k\in\mathbb{Z}$,
and greater than $|L|$. Then the final polymer amplitude, using step
3 mentioned before (for (\ref{eq:A-with-S}) and (\ref{eq:S-compact})),
is
\begin{equation}
A_{\textrm{poly}}(q_{f},q_{i};\alpha)=\sum_{M=|L|}^{\infty}A_{\textrm{poly}}(q_{f},q_{i};M;\alpha)=\sum_{k=\frac{|L|}{2}}^{\infty}\frac{1}{\left(2k\right)!}\left(-\frac{i\alpha}{2m}\frac{\hbar^{2}}{\lambda^{2}}\right)^{2k}e^{i\alpha\frac{\hbar^{2}}{m\lambda^{2}}}\begin{pmatrix}2k\\
k+\frac{L}{2}
\end{pmatrix}.\label{eq:A-poly-sum-M}
\end{equation}
Using a change of index, $k\rightarrow k+\frac{|L|}{2}$, in the above
yields
\begin{equation}
A_{\textrm{poly}}(q_{f},q_{i};\alpha)=\left(-1\right)^{L}i^{L}e^{i\alpha\frac{\hbar^{2}}{m\lambda^{2}}}J_{L}\left(\frac{\alpha}{m}\frac{\hbar^{2}}{\lambda^{2}}\right),\label{eq:A-poly-fin-free}
\end{equation}
with $J_{l}$ being the Bessel function of the first kind and we have
used the property $J_{n}(-x)=J_{-n}(x)=(-1)^{n}J_{n}(x)$. If we would
have assumed that $L$ is odd, the odd values of $M$ for which $M=2k+1$
and $M\geq|L|$ would have given rise to nonvanishing terms, but then
again, we would have gotten the same result. Now the full propagator
using the above, combined with (\ref{eq:Prop-Both}) and (\ref{eq:non-rel-pt-part}),
becomes
\begin{equation}
K_{\textrm{poly}}\left(q_{f},t_{f};q_{i},t_{i}\right)=i^{L}e^{-i\left[t_{f}-t_{i}\right]\frac{\hbar}{m\lambda^{2}}}J_{L}\left(\frac{\left[t_{f}-t_{i}\right]}{m}\frac{\hbar}{\lambda^{2}}\right),\label{eq:K-poly-nonrel-discrete-fin}
\end{equation}
This result coincides with earlier results based on Hamiltonian approach \citep{Hugo3}. We will now discuss the continuum limit of
this propagator.

\subsubsection{The continuum limit}\label{sec:free-contnm-lmt}

One can get the continuum limit of a polymer quantized theory by letting
$\lambda\rightarrow0$, namely taking the fundamental discrete scale
to be zero. In our case, we assume that while $\lambda\rightarrow0$
and $L\rightarrow\infty$, the spatial interval remains finite, $(q_{f}-q_{i})=\lambda L<\infty$.
To find the continuum limit, we first rewrite (\ref{eq:K-poly-nonrel-discrete-fin})
in the form
\begin{equation}
K_{\textrm{poly}}\left(q_{f},t_{f};q_{i},t_{i}\right)=i^{L}e^{-iL^{2}\Delta}J_{L}\left(L^{2}\Delta\right),\,\,\,\,\,\,\,\,\,\,\,\,\,\,\Delta=\frac{\hbar\left(t_{f}-t_{i}\right)}{m\left(q_{f}-q_{i}\right)^{2}}.
\end{equation}
Using an identity of the Bessel functions,
\begin{equation}
J_{L}(L\sec(\beta))\approx\frac{\exp\left[iL\left(\tan(\beta)-\beta\right)\right]}{\sqrt{2\pi L}\sqrt[4]{1-\sec^{2}(\beta)}},\,\,\,\,\,\,\,\,\,\,\,\,\,\,\,\sec(\beta)=L\Delta,
\end{equation}
in the above $K_{\textrm{poly}}\left(q_{f},t_{f};q_{i},t_{i}\right)$
and then Taylor expanding each term separately around $z:=\left[\sec(\beta)\right]^{-1}\rightarrow0$
(i.e. taking $\lambda\rightarrow0$) and keeping only the first terms
yields
\begin{equation}
K_{\textrm{poly}}\left(q_{f},t_{f};q_{i},t_{i}\right)\approx\lambda K_{0}\left(q_{f},t_{f};q_{i},t_{i}\right),\label{eq:Kpoly-K-free}
\end{equation}
where
\begin{equation}
K_{0}\left(q_{f},t_{f};q_{i},t_{i}\right)=\sqrt{\frac{m}{2\pi i\hbar\left(t_{f}-t_{i}\right)}}\exp\left[\frac{im\left(q_{f}-q_{i}\right)^{2}}{2\hbar\left(t_{f}-t_{i}\right)}\right],
\end{equation}
is the standard propagator of the free particle. Note that the polymer
propagator corresponds to a countable measure due to the presence
of $\lambda$ in (\ref{eq:Kpoly-K-free}). This permits us to go from
discrete sum over positions to a corresponding integral. More precisely, the presence of $\lambda$ in the right side of (\ref{eq:Kpoly-K-free}) is necessary
for following reason. To compute the evolution of an initial wave
equation, $\psi_{0}(q)$, we make use of the propagator and the measure
of the Hilbert space in the following way, 
\begin{equation}
\psi(q,t)=\sum_{q^{\prime}\in Lattice}K_{\text{Poly}}(q,t;q^{\prime},0)\psi_{0}(q)=\sum_{q^{\prime}\in Lattice}\lambda K_{0}(q,t;q^{\prime},0)\psi_{0}(q)+\mathcal{O}(\lambda^{2}).
\end{equation}
If we consider the limit $\lambda\rightarrow0$ in this
equation, we get
\begin{equation}
\sum_{q^{\prime}\in Lattice}K_{\text{Poly}}(q,t;q^{\prime},0)\psi_{0}(q)\rightarrow\int dq^{\prime}K_{0}(q,t;q^{\prime},0)\psi_{0}(q),
\end{equation}
which shows the presence of $\lambda$ allows us to go
from the discrete sum to the corresponding integral, $\sum_{q}\lambda\rightarrow\int dq$, cf. \cite{Creutz85}.

\subsection{Particle in a box}

This case is essentially the free particle with a specific boundary
condition. We implement this boundary condition by the method of images
used in optics as will be explained in detail in what follows. 

To find the polymer amplitude, we start by deriving the form of the
function $\bar{S}(q_{f},q_{i};M)$ which is the equivalent of the
function $S(q_{f},q_{i};M)$ in the case of the free particle. To
implement the boundary conditions, we maintain that
\begin{equation}
\bar{S}(q_{f},q_{i};M)\bigg|_{q_{f}=x^{\pm}}=0,
\end{equation}
where $x^{-}$ and $x^{+}$ are the positions of the left and right
walls of the box respectively. Following the method of images, we
propose that the first contribution to $\bar{S}(q_{f},q_{i};M)$ would
be $S(q_{f},q_{i};M)$ of the free particle. The other contributions
come from the implementation of the boundary conditions through images.

\begin{figure}
\includegraphics[scale=.6]{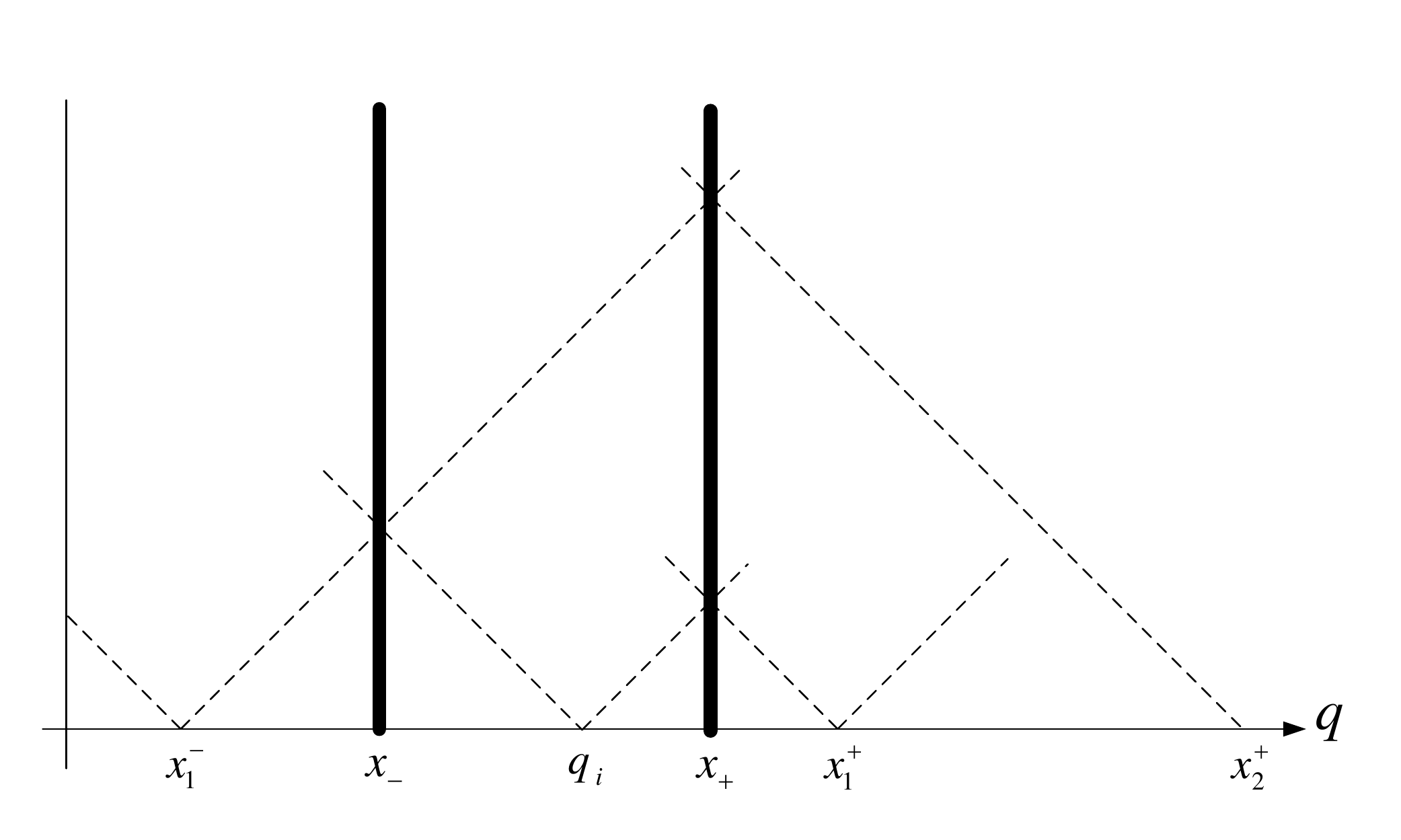}\caption{The particle in a box and the first three mirror images\label{fig:Box-mirror-points}}
\end{figure}

Now consider the image of the point $q_{i}$ with respect to the left
wall at $x^{-}$. We call the position of this image,  $x_{1}^{-}$ (see Fig. \ref{fig:Box-mirror-points}).
On the other hand, one can see from the properties of $S$ in (\ref{eq:S-prop-1})
and (\ref{eq:S-prop-2}) that $S(q_{i}+a,q_{i};M)=S(q_{i}-a,q_{i};M)$.
Using these, and noting that $q_{i}-x^{-}=x^{-}-x_{1}^{-}$, the boundary
condition at $x^{-}$, which is  $\bar{S}(x^{-},q_{i};M)=0$, is satisfied if
we assume 
\begin{equation}
\bar{S}(q_{f},q_{i};M)=S(q_{f},q_{i};M)-S(q_{f},x_{1}^{-};M).
\end{equation}
Now we use the same argument to implement the boundary condition at
$x^{+}$, $\bar{S}(x^{+},q_{i};M)=0$. Consider the images $x_{1}^{+}$
and $x_{2}^{+}$ of the point $q_{i}$ and $x_{1}^{-}$, respectively,
with respect to the right wall at $x^{+}$ (see Fig. \ref{fig:Box-mirror-points}).
In this case we have $x^{+}-q_{i}=x_{1}^{+}-x^{+}$ and $x^{+}-x_{1}^{-}=x_{2}^{+}-x^{+}$.
Then the boundary condition at $x^{+}$ is satisfied if we take
\begin{equation}
\bar{S}(q_{f},q_{i};M)=S(q_{f},q_{i};M)-S(q_{f},x_{1}^{-};M)-S(q_{f},x_{1}^{+};M)+S(q_{f},x_{2}^{+};M).
\end{equation}
Obviously these combination of functions does not satisfy the boundary
condition at $x^{-}$ and we have to again introduce images of $x_{1}^{+}$
and $x_{2}^{+}$ with respect to the left wall and repeat the same procedure again.
This process should be done indefinitely. To generalize this method,
we note that the position of images can be written as
\begin{equation}
x_{n}^{\pm}=2x^{\pm}-x_{n-1}^{\mp},\,\,\,\,\,\,\,\,\,\,\,\,\,\,n\in\mathbb{N},\,x_{0}^{\pm}=q_{i},
\end{equation}
and the contributions corresponding to even $n$ are positive while
the ones corresponding to the odd ones are negative. The above leads
to
\begin{equation}
x_{n}^{\pm}=\pm2L+x_{n-2}^{\pm},
\end{equation}
where $L=x^{+}-x^{-}$ is the length of the box. The solutions to
these difference equations for even and odd $n$ are
\begin{align}
x_{2k}^{\pm}= & \pm2kL+q_{i},\,\,\,\,\,\,\,\,\,\,\,\,\,n=2k,\,k\in\mathbb{N}\label{eq:x-pm-2k}\\
x_{2k-1}^{\pm}= & \pm2\left(k-1\right)L+2x^{\pm}-q_{i},\,\,\,\,\,\,\,\,\,\,\,\,\,n=2k-1,\,k\in\mathbb{N}.\label{eq:x-pm-2k-1}
\end{align}
From these one can see that
\begin{align}
x_{2k}^{-}\bigg|_{k\rightarrow-k}= & x_{2k}^{+},\label{eq:simpl-odd-even-1}\\
x_{2k-1}^{-}\bigg|_{k\rightarrow-k+1}= & x_{2k-1}^{+}.\label{eq:simpl-odd-even-2}
\end{align}
Now, the full contribution to $\bar{S}(q_{f},q_{i};M)$ comes from
the sum of three groups of functions: $S(q_{f},x_{2k}^{\pm};M)$ with
positive sign, $S(q_{f},x_{2k-1}^{\pm};M)$ with negative sign, and
$S(q_{f},q_{i};M)$ also with positive sign,
\begin{align}
\bar{S}(q_{f},q_{i};M)= & S(q_{f},q_{i};M)+\sum_{k=1}^{\infty}\left[S(q_{f},x_{2k}^{+};M)+S(q_{f},x_{2k}^{-};M)\right]-\sum_{k=1}^{\infty}\left[S(q_{f},x_{2k-1}^{+};M)+S(q_{f},x_{2k-1}^{-};M)\right].
\end{align}
Using (\ref{eq:simpl-odd-even-1}) and (\ref{eq:simpl-odd-even-2}),
and letting $k\in\mathbb{Z}$, the above function can be expressed more
concisely as
\begin{align}
\bar{S}(q_{f},q_{i};M)= & \sum_{k=-\infty}^{\infty}\left[S(q_{f},x_{2k}^{+};M)-S(q_{f},x_{2k-1}^{+};M)\right]\nonumber \\
= & \sum_{k=-\infty}^{\infty}\left[S(q_{f},2kL+q_{i};M)-S(q_{f},2kL+2x^{-}-q_{i};M)\right]
\end{align}
where in the second line we have used (\ref{eq:x-pm-2k}) and (\ref{eq:x-pm-2k-1}).

Finally using this function together with (\ref{eq:S-compact}) in
(\ref{eq:A-with-S}) and summing over $M$ (or using (\ref{eq:A-poly-sum-M})),
and then using the resulting expression together with (\ref{eq:non-rel-pt-part})
in (\ref{eq:Prop-Both}), yields
\begin{equation}
K_{\textrm{poly}}^{\textrm{Box}}\left(q_{f},t_{f};q_{i},t_{i}\right)=\sum_{k=-\infty}^{\infty}\left[K_{\textrm{poly}}\left(q_{f},t_{f};q_{i}+2kL,t_{i}\right)-K_{\textrm{poly}}\left(q_{f},t_{f};-q_{i}+2kL,t_{i}\right)\right],\label{eq:Full-box-prop}
\end{equation}
where $K_{\textrm{poly}}\left(q_{f},t_{f};q_{i},t_{i}\right)$ is
the polymer propagator of the free particle derived in the previous
section, and we have chosen $x^{-}=0$. This is precisely the result
that has been obtained by spectral methods in \citep{Hugo3}.

\subsubsection{The continuum limit}

Due to the form of $K_{\textrm{poly}}^{\textrm{Box}}\left(q_{f},t_{f};q_{i},t_{i}\right)$,
its continuum limit simply results from the continuum limit of the
propagator of the free particle,
\begin{align}
K_{\textrm{poly}}^{\textrm{Box}}\approx & \lambda\sum_{k=-\infty}^{\infty}\left[K_{0}\left(q_{f},t_{f};q_{i}+2kL,t_{i}\right)-K_{0}\left(q_{f},t_{f};-q_{i}+2kL,t_{i}\right)\right]\nonumber \\
\approx & \lambda K^{\textrm{Box}}\left(q_{f},t_{f};q_{i},t_{i}\right),
\end{align}
with $K^{\textrm{Box}}\left(q_{f},t_{f};q_{i},t_{i}\right)$ being
the usual propagator of the particle in a box. The presence of $\lambda$ here was discussed earlier at the end of Sect. \ref{sec:free-contnm-lmt}.

\section{Relativistic particle\label{sec:rel-partcl}}

For this case we assume that the system is four dimensional. The deparametrized
constraint is $\mathcal{C}=\pi^{2}-h(q,p)=p_{t}^{2}-\sum_{j=1}^{3}\left(p^{j}\right)^{2}-m^{2}$
with $\pi^{2}=p_{t}^{2}$, which is the four dimensional generalization
of (\ref{eq:rel-deparametr}) and (\ref{eq:rel-constraint}). The
representation of this constraint on the Hilbert space can be written
as
\begin{equation}
\hat{\mathcal{C}}=\hat{p}_{t}^{2}-\sum_{j=1}^{3}\widehat{\left(p^{j}\right)_{\lambda}^{2}}-m^{2},
\end{equation}
where  $\hat{p}_{t}^{2}$ is the usual Schr\"{o}dinger representation
of $p_{t}$ while $\widehat{\left(p^{j}\right)_{\lambda}^{2}}$ is
\begin{equation}
\widehat{\left(p^{j}\right)_{\lambda}^{2}}=\frac{\hbar^{2}}{\lambda^{2}}\left(2-\hat{V}_{\lambda}^{j}-\hat{V}_{-\lambda}^{j}\right),
\end{equation}
which is the polymer representation of $\left(p^{j}\right)^{2}$ similar
to (\ref{eq:P2-poly}). The full polymer propagator from (\ref{eq:Prop-Both})
becomes
\begin{equation}
K_{\textrm{poly}}\left(q_{f},t_{f};q_{i},t_{i}\right)=\int_{\mathbb{R}}d\alpha\,e^{-i\alpha m^{2}}\langle t_{f}|e^{i\alpha\hat{p}_{t}^{2}}|t_{i}\rangle\prod_{j=1}^{3}\langle q_{f}^{j}|e^{-i\alpha\widehat{\left(p^{j}\right)_{\lambda}^{2}}}|q_{i}^{j}\rangle,\label{eq:K-poly-relat-initial}
\end{equation}
where the superscript $j$ on $q$, corresponds to the three spatial
dimensions. The amplitude of the relativistic momentum of the physical time,
(\ref{eq:A-pi-Both}), turns out to be
\begin{equation}
A_{\pi}(t_{f},t_{i};\alpha)=\langle t_{f}|e^{i\alpha\hat{\pi}^{2}}|t_{i}\rangle=\int_{\mathbb{R}}dp_{t}e^{i\alpha p_{t}^{2}}e^{\frac{i}{\hbar}p_{t}\left(t_{f}-t_{i}\right)}=\sqrt{\frac{i\pi}{\alpha}}e^{-\frac{i}{4\hbar^{2}\alpha}\left(t_{f}-t_{i}\right)^{2}}.\label{eq:A-pi-relat-fin}
\end{equation}
Due to the similarity of $\hat{h}$ in both relativistic and nonrelativistic
particle cases (see (\ref{eq:non-rel-constraint}), (\ref{eq:rel-constraint})
and (\ref{eq:Prop-Both})), the relativistic polymer amplitudes can
be computed, to a great extent,  in the same manner as in section \ref{subsec:Free-particle}.
The only difference is that we need to make a simple change, $\lambda^{2}\rightarrow\frac{\lambda^{2}}{-2m},$
in the result in (\ref{eq:A-poly-fin-free}), and keep in mind that
now there are three lengths $L^{j}=\frac{q_{f}^{j}-q_{i}^{j}}{v}$
instead of just one length $L$. Therefore we will get
\begin{equation}
A_{\textrm{poly}}(q_{f},q_{i};\alpha)=\langle q_{f}^{j}|e^{-i\alpha\widehat{\left(p^{j}\right)_{\lambda}^{2}}}|q_{i}^{j}\rangle=\left(-1\right)^{L^{j}}i^{L^{j}}e^{-i\alpha\frac{2\hbar^{2}}{\lambda^{2}}}J_{L^{j}}\left(-2\alpha\frac{\hbar^{2}}{\lambda^{2}}\right),\,\,\,\,\,\,\,\,\,\,\,\,j=1,2,3.\label{eq:A-poly-relat}
\end{equation}
Putting (\ref{eq:K-poly-relat-initial}), (\ref{eq:A-pi-relat-fin})
and (\ref{eq:A-poly-relat}) together yields
\begin{equation}
K_{\textrm{poly}}^{\textrm{Rel}}\left(q_{f},t_{f};q_{i},t_{i}\right)=\left(\prod_{j=1}^{3}i^{L^{j}}\right)\int_{\mathbb{R}}d\alpha\,\sqrt{\frac{i\pi}{\alpha}}e^{-i\alpha m^{2}}e^{-\frac{i}{4\hbar^{2}\alpha}\left(t_{f}-t_{i}\right)^{2}}e^{-i\alpha\frac{6\hbar^{2}}{\lambda^{2}}}\left[\prod_{j=1}^{3}J_{L^{j}}\left(2\alpha\frac{\hbar^{2}}{\lambda^{2}}\right)\right],\label{eq:Full-prop-relat}
\end{equation}
which is the polymer propagator of the free relativistic particle.

\subsubsection{The continuum limit}

The Bessel function in this case becomes
\begin{equation}
J_{L^{j}}\left(2\alpha\frac{\hbar^{2}}{\lambda^{2}}\right)=J_{L^{j}}\left(2\alpha L^{j2}\frac{\hbar^{2}}{L^{j2}\lambda^{2}}\right)=J_{L^{j}}\left(L^{j2}\frac{2\alpha\hbar^{2}}{\left(q_{f}-q_{i}\right)^{2}}\right)=J_{L^{j}}\left(L^{j2}\Delta\right),
\end{equation}
where we have used $\Delta=\frac{2\alpha\hbar^{2}}{\left(q_{f}^{j}-q_{i}^{j}\right)^{2}}$,
and $L^{j}=\frac{q_{f}^{j}-q_{i}^{j}}{\lambda}$ is the length of
the $j$'th dimension of the particle's motion. By applying the same
approximation method used in section (\ref{subsec:Free-particle})
one gets (for the limit $\lambda\rightarrow0$)
\begin{equation}
J_{L}\left(L^{j2}\Delta\right)\approx\frac{1}{\sqrt{2\pi L^{j}}}\frac{\sqrt{z}}{\sqrt{i}}\left(-i\right)^{L^{j}}\exp\left[iL^{j}z^{-1}+\frac{1}{2}iL^{j}z\right],\,\,\,\,\,\,\,\,\,\,\,\sec(\beta)=L^{j}\Delta=z^{-1}.
\end{equation}
Substituting back for the values of $L^{j}$ and $z$, yields
\begin{equation}
J_{L^{j}}\left(2\alpha\frac{\hbar^{2}}{\lambda^{2}}\right)=\frac{\lambda}{\sqrt{4\pi i\alpha\hbar^{2}}}\left(-i\right)^{L^{j}}\exp\left[2i\alpha\frac{\hbar^{2}}{\lambda^{2}}+i\frac{\left(q_{f}^{j}-q_{i}^{j}\right)^{2}}{4\alpha\hbar^{2}}\right].
\end{equation}
Finally using it in the propagator (\ref{eq:Full-prop-relat}), one
arrives at
\begin{equation}
K_{\textrm{poly}}^{\textrm{Rel}}\left(q_{f},t_{f};q_{i},t_{i}\right)=\frac{\lambda^{3}}{\hbar^{3}}\frac{1}{8i\pi}\int_{\mathbb{R}}d\alpha\,\frac{e^{-i\alpha m^{2}}}{\alpha^{2}}\exp\left\{ \frac{i}{4\alpha\hbar^{2}}\left(\Delta S\right)^{2}\right\} =\frac{\lambda^{3}}{\hbar^{3}}\frac{1}{8i\pi}\int_{\mathbb{R}}d\alpha\,\frac{e^{\frac{i}{4\alpha\hbar^{2}}\left[\left(\Delta S\right)^{2}-4\alpha^{2}\hbar^{2}m^{2}\right]}}{\alpha^{2}},\label{eq:K-poly-rel-approx-contin}
\end{equation}
where
\begin{equation}
\left(\Delta S\right)^{2}=-\left(t_{f}-t_{i}\right)^{2}+\sum_{j}\left(q_{f}^{j}-q_{i}^{j}\right)^{2}=\eta_{\mu\nu}\left(x_{f}^{\mu}-x_{i}^{\mu}\right)\left(x_{f}^{\nu}-x_{i}^{\nu}\right)
\end{equation}
is the Lorentz invariant spacetime interval between initial and final
points of the particle's motion in flat spacetime. This can be written
as
\begin{equation}
K_{\textrm{poly}}^{\textrm{Rel}}\left(q_{f},t_{f};q_{i},t_{i}\right)=\lambda^{3}K^{\textrm{Rel}},
\end{equation}
where $K^{\textrm{Rel}}$ is the well-known propagator of the nonpolymer
relativistic particle. Again, the presence of $\lambda^3$ is due to the reasons discussed at the end of Sect. \ref{sec:free-contnm-lmt}.

One can make an interesting observation using (\ref{eq:K-poly-rel-approx-contin}):
assuming that the particle is a massless one that moves on a null
curve in spacetime, one has $\left(\Delta S\right)^{2}=0$ and the
propagator becomes
\begin{equation}
K_{\textrm{poly}}^{\textrm{Rel}}\left(q_{f},t_{f};q_{i},t_{i}\right)=\frac{\lambda^{3}}{\hbar^{3}}\frac{1}{8i\pi}\lim_{m\rightarrow0}\int_{\mathbb{R}}d\alpha\,\frac{e^{-i\alpha m^{2}}}{\alpha^{2}}.
\end{equation}
The integral can be computed by analytic continuation and by using
the incomplete Gamma function $\Gamma(a,z)$. This will yield
\begin{align}
K_{\textrm{poly}}^{\textrm{Rel}}\left(q_{f},t_{f};q_{i},t_{i}\right)= & \frac{\lambda^{3}}{\hbar^{3}}\frac{1}{8i\pi}\lim_{m\rightarrow0}\left[-\frac{1}{\alpha}e^{-i\alpha m^{2}}+im^{2}\Gamma\left(0,i\alpha m^{2}\right)\right]=-\frac{\lambda^{3}}{\hbar^{3}}\frac{1}{8i\pi\alpha}.
\end{align}
We can recognize the usual Feynman propagator of a massless particle  on the right hand side of the above equation. Once again the appearance of the factor $\lambda^3$ can be understood as in the previous cases.

\section{Discussion}\label{sec:discuss}

It is a general expectation  that modifying the underlying structure of spacetime can alleviate the difficulties posed by the spacetime singularities of General Relativity and the ultraviolet divergences of field theory. In particular proposals like Loop Quantum Gravity which lead to a discrete  geometry  may play an important role in this regard.  Indeed some cosmological  models as well as black hole interiors have been  shown to avoid  the classical singularities when they are described using LQG methods.  Moreover, along the same lines, field theories  improve their high energy behavior.  Now  there are two approaches  that have been developed to implement loop quantization, a Hamiltonian one and a path integral  version, better known as spin foam models. A recent proposal  was put forward to connect both approaches  in the case of cosmological models that is based on a vertex expansion  of a transition amplitude. However such an amplitude has not been calculated completely in an explicit way in spite of the fact that the models considered are soluble. 

In this work we have applied a previous proposal to connect the Hamiltonian and path integral polymer quantizations to  three mechanical systems in order to
calculate their explicit analytic polymer propagators. These systems
are the nonrelativistic particle, both free (\ref{eq:K-poly-nonrel-discrete-fin})
and in a box (\ref{eq:Full-box-prop}), and the relativistic particle
(\ref{eq:Full-prop-relat}). Although the first two had been computed
before through spectral methods \citep{Hugo3}, they are regained
here using path integration.
The third one is a completely new result. All three are shown here
to reduce to the known forms, a la Schr\"{o}dinger, when the
polymer scale parameter is small. 

After giving a brief introduction to polymer representation, constrained
systems and their deparametrization, and group averaging technique, we
started by considering the deparametrized form of the classical canonical
description of a free particle %
\begin{comment}
in a position dependent potential
\end{comment}
{} where there is only one associated constraint present. Next we set
up the polymer quantization of this system for the two independent
degrees of freedom, time and position. The former is quantized a la
Schr\"{o}dinger while the latter is subject to the polymer quantization,
thus differing in their Hilbert spaces, as well as the commutator
algebra to be fulfilled, as can be seen from (\ref{eq:Algebra-t,q}).
This is similar to what is actually done in the cosmological model
\citep{AshtekarPI} where the clock  variable is taken to be a
scalar field quantized in the standard way, whereas the gravitational
degrees of freedom are treated using polymer quantization.

Thanks to $t$ and $q$ commuting, the generic
form of the polymer propagator can be expressed as the product of a polymer amplitude
corresponding to $q$ and a Schr\"{o}dinger one associated to $t$.
As for the polymer  contribution the splitting into small propagators which amounts to a sum over histories can be given the form of a sum over a number of transitions that involve a change in position of the particle. Such resummation is just the analogue of the vertex expansion of the spin foam models that was identified in \cite{AshtekarPI,A2}. In contrast to the latter, however, for our mechanical models we have been able to compute the sums explicitly and hence derive an analytic propagator for each case. 

The calculations are remarkably simplified in the case of a free non
relativistic particle. For a particle in a box, the boundary walls
are implemented through combinations of free particle contributions
via the image method. Moreover, this deparametrized approach turns
out to be very convenient for the relativistic particle due to the
fact that the time and position variables decouple.

Our explicit results for the polymer propagators lend support to the proposal to connect the Hamiltonian and the path integral loop quantizations. 

The present work can be extended in several directions. One possibility is 
to pursue the path integral formulation of the polymer harmonic 
oscillator which will have various applications in polymer field theory \citep{Hossain}. Our analysis may also suggest some approximations in the case of the cosmological models for which the explicit propagator has not been obtained, in particular in regard to the resummmation of the amplitude in terms of a vertex expansion. Additionally, there are indications that polymer field theory could lead to violation of Lorentz invariance symmetry \citep{Hossain}, although this has been shown to be phenomenologically constrained \cite{PRL11606}. More work is needed  to settle this issue, perhaps using the path integral method, since these types of effects could be an artifact of the truncation used to derive the dispersion relation of the model  \cite{PRD71084} (see also \cite{Kajuri:2014kva}).

\begin{acknowledgments}
We would
like to acknowledge the  support of CONACyT grant number
237351: Implicaciones F\'{i}sicas de la Estructura del Espaciotiempo.
S.R. would like to acknowledge the support of the PROMEP postdoctoral
fellowship (through UAM-I), and the grant from Sistema Nacional de
Investigadores of CONACyT. 
\end{acknowledgments}

\bibliography{main}

\end{document}